\documentclass[journal,oneside]{IEEEtran}
\usepackage{mathtools}

\usepackage{graphicx}
\usepackage{lettrine}
\usepackage{booktabs}
\usepackage{amsmath}
\usepackage{enumerate}
\usepackage{bm}
\usepackage{graphicx}
\usepackage{epstopdf}
\usepackage{amsmath}
\usepackage{array}
\usepackage{amssymb}
\usepackage[numbers,sort&compress]{natbib}
\usepackage{placeins}
\usepackage{stfloats}
\usepackage{epstopdf}

\usepackage{xcolor}
\usepackage{soul}

\newcolumntype{P}[1]{>{\raggedright\arraybackslash}p{#1}}
\newcommand{\tabincell}[2]{\begin{tabular}{@{}#1@{}}#2\end{tabular}}

\ifCLASSINFOpdf
\else
\fi
  \usepackage[caption=false,font=footnotesize]{subfig}
\usepackage{tikz}
\usepackage{textcomp}

\newcommand\copyrighttext{%
  \footnotesize \textcopyright 2019 IEEE. Personal use of this material is permitted.
  Permission from IEEE must be obtained for all other uses, in any current or future 
  media, including reprinting/republishing this material for advertising or promotional 
  purposes, creating new collective works, for resale or redistribution to servers or 
  lists, or reuse of any copyrighted component of this work in other works. 
  DOI: 10.1109/TPWRS.2019.2953225}
\newcommand\copyrightnotice{%
\begin{tikzpicture}[remember picture,overlay]
\node[anchor=south,yshift=5pt] at (current page.south) {\fbox{\parbox{\dimexpr\textwidth-\fboxsep-\fboxrule\relax}{\copyrighttext}}};
\end{tikzpicture}%
}

\begin{document}
%
\title{Scale- and Context-Aware Convolutional Non-intrusive Load Monitoring}

\author{
Kunjin Chen, Yu Zhang, \emph{Member, IEEE}, Qin Wang, Jun Hu, \emph{Member, IEEE}, Hang Fan, and Jinliang He, \emph{Fellow, IEEE} 

\thanks{
K. J. Chen, J. Hu, H. Fan, and J. L. He are with the State Key Lab of Power Systems, Department of Electrical Engineering, Tsinghua University, Beijing 100084, P. R. of China. 

Y. Zhang is with the Department of Electrical and Computer Engineering, University of California, Santa Cruz, Santa Cruz, CA 95064, USA. 

Q. Wang is with the intelligent maintenance systems group, ETH Z\"urich, 8092 Z\"urich, Switzerland.

(Corresponding author email: hejl@tsinghua.edu.cn).
} 
}


%

\maketitle

\copyrightnotice
\vspace{-10pt}

\begin{abstract}
Non-intrusive load monitoring addresses the challenging task of decomposing the aggregate signal of a household's electricity consumption into appliance-level data without installing dedicated meters. By detecting load malfunction and recommending energy reduction programs, cost-effective non-intrusive load monitoring provides intelligent demand-side management for utilities and end users. In this paper, we boost the accuracy of energy disaggregation with a novel neural network structure named scale- and context-aware network, which exploits multi-scale features and contextual information. Specifically, we develop a multi-branch architecture with multiple receptive field sizes and branch-wise gates that connect the branches in the sub-networks. We build a self-attention module to facilitate the integration of global context, and we incorporate an adversarial loss and on-state augmentation to further improve the model's performance. Extensive simulation results tested on open datasets corroborate the merits of the proposed approach, which significantly outperforms state-of-the-art methods.
\end{abstract}

\smallskip


%
\IEEEpeerreviewmaketitle

\section{Introduction}

Non-intrusive load monitoring (NILM) is the task of estimating the power demand of a specific appliance from the aggregate consumption of a household measured by a single meter \cite{hart1992nonintrusive}. As the task requires breaking down the total energy consumed by multiple appliances into appliance-level energy consumption records, NILM is synonymous with the phrase ``energy disaggregation'' \cite{shin2018subtask}. A direct benefit of NILM is that energy end-users can acquire appliance-level consumption feedbacks and optimize their energy consumption behaviours accordingly. It is estimated that up to 12\% residential energy saving can be achieved by providing appliance-level feedback \cite{ehrhardt2010advanced}. NILM benefits consumers, the research community and utilities in domains including residential and commercial energy use, appliance innovation, energy efficient marketing and program evaluation \cite{armel2013disaggregation}.

The approaches for NILM can be generally divided into supervised methods and unsupervised methods \cite{bonfigli2015unsupervised}. In the supervised setting, the power consumptions of individual appliances are collected and can be used to train the models. For unsupervised methods, however, only the aggregate power comsumption data can be used. Approaches for unsupervised NILM include hidden Markov models (HMM)\cite{parson2012non, parson2014unsupervised}, factorial hidden Markov models (FHMM) \cite{kolter2012approximate, NIPS2016_6534} and methods based event detection and clustering \cite{gonccalves2011unsupervised, zhao2016training}. Comprehensive reviews of unsupervised NILM approaches can be found in \cite{bonfigli2015unsupervised, zhuang2018overview}. 

With the development of deep neural networks, various neural network-based supervised NILM approaches have been proposed \cite{mauch2015new, kelly2015neural}. A substantial progress has been made recently thanks to convolutional neural networks (CNN) \cite{zhang2018sequence, shin2018subtask}. For the task of NILM, the power consumption patterns of different appliances generally have varied scales. The aggregate consumption of multiple appliances is prone to have more complicated shapes, hence requiring the ability to deal with scale variation. In addition to local information within a small time range, it is also important to consider the context dependencies of consumption patterns as energy consumption behaviours contain higher-level semantics (e.g., the dryer works after the washer, and one may turn on the microwave multiple times until cooking is finished). However, existing CNN-based models fail to exploit those aspects, which yield high rates of false positive/negative errors in the disaggregation results. In light of this, we propose a scale- and context-aware network (SCANet) structure to incorporate the above-mentioned ideas. In this paper, we compare the performance of SCANet with state-of-the-art models and conduct empirical analyses on the advantages of the proposed structure. The contributions of this work are twofold:
\begin{itemize}
\item A scale- and context-aware CNN structure is designed for the task of NILM, which greatly improves the disaggregation results for multiple appliances. 
\item We show that adding adversarial loss or on-state augmentation can help the model produce more accurate results and increase generalizability.
\end{itemize}

The organization of the rest of the paper is as follows: we introduce the related work of this study in Section II. The modules for scale and context awareness, the adversarial loss and the on-state augmentation are described in detail in Section III. The effectiveness of the proposed SCANet model is validated in Section IV with extensive comparisons and visualizations. An additional experiment setting that uses partial ground truth is also introduced and implemented. Finally, Section V concludes the paper and points out some future works. 

\section{Related Work}

\subsubsection{Neural Non-intrusive Load Monitoring}

The application of neural networks in NILM started with recurrent neural networks (RNN), CNN, and denoising auto-encoders (DAE) with relatively simple structures \cite{mauch2015new,kelly2015neural}. Various CNN models have been proposed thanks to the flexibility of CNN structures, such as sequence-to-point, sequence-to-sequence, and fully convolutional models \cite{zhang2018sequence,chen2018convolutional,brewitt2018non}. 

The integration of domain knowledge further enriches the design of CNN architectures. An on/off state classification sub-network can be added in parallel to the regression sub-network so that the model can learn from on/off state information directly \cite{shin2018subtask,murray2018transferability}. The work in this paper adopts the structure of subtask gated network (SGN) \cite{shin2018subtask} as a starting point.

\subsubsection{Multi-scale Features in CNNs}

CNNs are widely used in computer vision tasks including object detection and semantic segmentation, for which capturing multi-scale information is of crucial significance. When features of various scales exist in a CNN structure, these features can be combined by upsampling higher layers \cite{hariharan2015hypercolumns} or adopting different sampling strategies (e.g., use either max pooling or deconvolution) for different layers \cite{kong2016hypernet}. The association of multi-scale features can also be achieved by building pyramid-like network structures such as U-NET \cite{ronneberger2015u}, FPN \cite{lin2017feature}, and PANet \cite{liu2018path}. While U-NET concatenates low-level features and upsampled high-level features using skip-connections in a sequential manner and uses the last layer for prediction, features of multiple layers are used by FPN to produce predictions of various scales.

Another way to create features of multiple scales is to use dilated convolutions \cite{yu2015multi}, which is adopted by TridentNet \cite{li2019scale} to generate multi-scale features in several parallel branches with different dilation rates. Scale awareness is obtained by training the branches separately with objects within certain scale ranges. In this work, we use a multi-branch structure similar to that of TridentNet. Unlike TridentNet, however, we use gating signals generated by branches in the on/off state classification network to selectively keep feature maps in the regression sub-network, which facilitates scale awareness.

\subsubsection{Self-attention Mechanism}

The attention mechanism is useful when additional information can be provided by global context \cite{bahdanau2014neural}. For self-attention, the output value at a position in a sequence is calculated by attending to all positions in the sequence \cite{zhang2018self}. Applications including machine translation and video classification greatly benefit from the usage of self-attention \cite{vaswani2017attention,wang2018non}. In this work, we adopt the self-attention module proposed by Zhang \emph{et al.} \cite{zhang2018self}. 

\subsubsection{Generative Adversarial Networks}

Generative adversarial networks (GAN) are a family of generative models that are capable of generating realistic data \cite{goodfellow2014generative}, and different extensions of GANs have been applied to tasks including image-to-image translation \cite{zhu2017unpaired}, image inpainting \cite{yu2018generative}, text-to-image synthesis \cite{reed2016generative}, music generation \cite{yang2017midinet}, etc. Other works focus on stabilizing the training of GANs and improve the quality of generated samples \cite{arjovsky2017wasserstein, gulrajani2017improved, zhao2017energy, takeru2018spectral}. Applying GANs to NILM is a relatively new idea \cite{bao2018enhancing}, where a disaggregator is used to produce latent representations for a specific appliance, followed by a generator that produces the load sequence of the appliance. Different from \cite{bao2018enhancing}, we directly formulate the generator as a mapping from the aggregate consumption to the appliance-level consumption without producing the latent representations.

\section{Proposed Model and Training Techniques}

In this section, we first formally define the NILM task considered in the paper. We then briefly introduce existing CNN-based models and elaborate on the building blocks of SCANet. Techniques that can facilitate the training of the model are also introduced.

\subsection{Problem Formulation}

\begin{figure}[!tb]
\centering
\includegraphics[width=7cm]{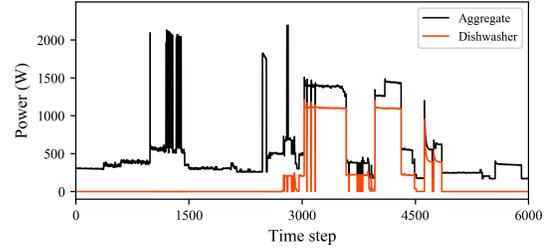}
\caption{An illustration of the NILM task: recovering the power consumption signal of a dishwasher from the aggregate consumption profile.}
\label{task}
\end{figure}

Consider a household with a given aggregate power consumption signal $\tilde{\mathbf{x}} = (x_1, \cdots, x_T)$. Let $\tilde{\mathbf{y}}^i = (y^i_1, \cdots, y^i_T)$ and $\tilde{\mathbf{u}} = (u_1, \cdots, u_T)$ denote the power consumption of the $i$th appliance being considered and the total consumption of all remaining appliances, respectively. Then, we have $x_t = \sum_{i=1}^{N_a}{y^i_t} + u_t + \epsilon_t$, where $N_a$ is the number of appliances and $\epsilon_t$ is the additive noise. Given the aggregate signal $\tilde{\mathbf{x}}$, the task of NILM is to recover the power consumption sequences $\tilde{\mathbf{y}}^1, \tilde{\mathbf{y}}^2, \cdots, \tilde{\mathbf{y}}^{N_a}$ of the appliances under consideration \cite{shin2018subtask}. An illustration of the task is provided in Fig. \ref{task}.

\subsection{Model Design}

Estimating the power consumption sequence of an appliance with length $s$ using the aggregate consumption signal corresponding to the same time window is difficult as contextual information outside the window is not considered. Thus, it is suggested to add windows of length $w$ to both sides for the input aggregate sequence \cite{shin2018subtask,chen2018convolutional}. Specifically, with input sequence $\mathbf{x}_{t} \mathrel{\mathop:}= (x_{t-w}, \cdots, x_{t+s+w-1})$, $\mathbf{y}^i_{t} \mathrel{\mathop:}= (y^i_t, \cdots, y^i_{t+s-1})$ is the output sequence for the $i$th appliance.

It is straightforward to formulate sequence-to-sequence neural network models with stacked convolutional layers and fully-connected (FC) layers for the task $\hat{\mathbf{y}}^i_t = f(\mathbf{x}_t)$, where $\hat{\mathbf{y}}^i_t$ is the predicted sequence \cite{zhang2018sequence}. In order to exploit the on/off state information, two sub-networks, namely, $f_{\mathrm{power}}:\mathbb{R}^{s+2w}_+ \rightarrow \mathbb{R}^s_+$ and $f_{\mathrm{on}}:\mathbb{R}^{s+2w}_+ \rightarrow [0,1]^s$, are formulated \cite{shin2018subtask}. An auxiliary sequence $\mathbf{o}^i_{t} \mathrel{\mathop:}= (o^i_t, \cdots, o^i_{t+s-1}) \in \left\{ 0, 1 \right\}^s$ representing the on/off state of the $i$th appliance is added and the predicted sequence of on-state probability is given as $\hat{\mathbf{o}}^i_t = f_{\mathrm{on}}(\mathbf{x}_t)$. Hence, the final output of the model is 
\begin{equation}
\hat{\mathbf{y}}^i_t = f^i_{\mathrm{output}}(\mathbf{x}_t) = f^i_{\mathrm{on}}(\mathbf{x}_t) \odot f^i_{\mathrm{power}}(\mathbf{x}_t),
\end{equation}
where $\odot$ is the element-wise multiplication. For simplicity, we omit the superscript $i$ and subscript $t$ hereafter.

\begin{figure}[!tb]
\centering
\includegraphics[width=7cm]{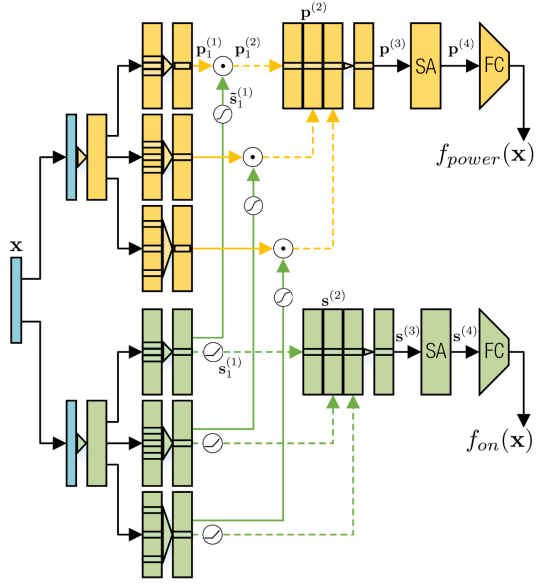}
\caption{The proposed scale- and context-aware structure.}
\label{main_structure}
\end{figure}

The structure of the two sub-networks proposed in \cite{zhang2018sequence, shin2018subtask} is illustrated in \mbox{Fig. \ref{subtask_structure}}. We build our model featuring scale and context awareness based on this structure (see Fig. \ref{main_structure}). The additional components are added based on two observations of existing works: first, the convolutional layers are unable to explicitly extract features with different time scales, and second, the features of the convolutional layers for a given time step are produced based on neighbouring input values without referring to the context. The details of the scare and context awareness modules are elaborated as follows:

\begin{figure}[!tb]
\centering
\includegraphics[width=5cm]{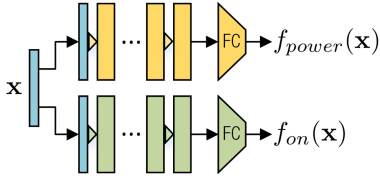}
\caption{The CNN structure with two sub-networks proposed in \protect\cite{shin2018subtask}.}
\label{subtask_structure}
\end{figure}

\subsubsection{Scale-aware Feature Extraction}

\begin{figure}[!tb]
\centering
\includegraphics[width=8.5cm]{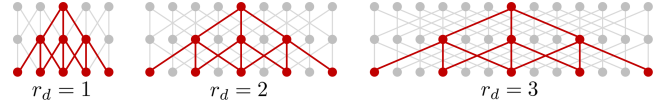}
\caption{An illustration of dilated convolution with different dilation rates.}
\label{dilated_conv}
\end{figure}

The scale awareness of SCANet is obtained by adding parallel branches with different dilation rates to the original network and connecting the branches in the two sub-networks by a simple gating mechanism, which allows the regression network keep only the most important feature maps at different scales. An illustration of dilated convolution with different dilation rates ($r_d=1,2,3$) is shown in Fig. \ref{dilated_conv}. With the same number of layers and parameters, a larger $r_d$ allows the output nodes respond to wider time ranges at the input. Thus, the outputs at the branches with different $r_d$ will reflect elements (e.g., shapes or edges) of different time scales at the input. At the same time, an element at the input will affect more output nodes when a larger $r_d$ is used.

Let $\left( \mathbf{p}^{(1)}_1, \mathbf{p}^{(1)}_2, \mathbf{p}^{(1)}_3 \right)$ be the outputs of the branches in $f_{\mathrm{power}}$ and let $\left( \tilde{\mathbf{s}}^{(1)}_1, \tilde{\mathbf{s}}^{(1)}_2, \tilde{\mathbf{s}}^{(1)}_3 \right)$ be the outputs of the branches in $f_{\mathrm{on}}$ with the sigmoid activation function. Then, the gating mechanism associating the two sub-networks is given by
\begin{equation}
\mathbf{p}^{(2)}_j = \mathbf{p}^{(1)}_j \odot \tilde{\mathbf{s}}^{(1)}_j, \; j=1,2,3.
\end{equation}
As the gating operation is separately performed for each dilation rate, a rich combination of features at different time scales can be achieved. We then concatenate $\left( \mathbf{p}^{(2)}_1, \mathbf{p}^{(2)}_2, \mathbf{p}^{(2)}_3 \right)$ and $\left( \mathbf{s}^{(1)}_1, \mathbf{s}^{(1)}_2, \mathbf{s}^{(1)}_3 \right)$ and obtain $\mathbf{p}^{(2)}$ and $\mathbf{s}^{(2)}$ (note that $\mathbf{s}^{(2)}$ contains features with the rectified linear unit (ReLU) activation function instead of the sigmoid function). Both $\mathbf{p}^{(2)}$ and $\mathbf{s}^{(2)}$ are processed by a convolutional layer with a kernel size of 1, yielding $\mathbf{p}^{(3)}$ and $\mathbf{s}^{(3)}$, the inputs to the self-attention modules.

\subsubsection{Context-aware Feature Integration}

The integration of contextual information is achieved by the self-attention module, which takes an input $\mathbf{z} \in \mathbb{R}^{C \times L}$ with $L$ time steps and $C$ channels. The module learns an additional feature map $\mathbf{r} \in \mathbb{R}^{C \times L}$ whose values at each time step are obtained by attending to all the time steps in $\mathbf{z}$. We first map the input $\mathbf{z}$ with $g(\mathbf{z}) = \mathbf{W}_g \mathbf{z}$ and $h(\mathbf{z}) = \mathbf{W}_h \mathbf{z}$, and an entry $a_{j,i}$ in the attention matrix $\mathbf{A}$ is calculated as
\begin{equation}
a_{j,i} = \frac{\exp(\tilde{a}_{i,j})}{\sum^L_{l=1}{\exp(\tilde{a}_{l,j})}}, \; {\rm where} \; \tilde{a}_{i,j} = \left[ g(\mathbf{z})^\top h(\mathbf{z}) \right]_{i,j}.
\end{equation}
The additional feature map $\mathbf{r}$ is then calculated by
\begin{equation}
\mathbf{r} = d(\mathbf{z})\mathbf{A}, \; {\rm where} \; d(\mathbf{z}) = \mathbf{W}_d \mathbf{z}.
\end{equation}
Note that $a_{j,i}$ is the attention assigned to the $i$th time step when the response of the $j$th time step is being calculated. The output of the self-attention module is defined as $\mathbf{z} + \gamma\mathbf{r}$, where $\gamma$ is initialized as 0 and updated when the model is trained, so that the model can rely on the local context at first and gradually learn to pick up the dependencies in the global context \cite{zhang2018self}. Specifically, weight matrices $\mathbf{W}_g \in \mathbb{R}^{\bar{C}\times C}$, $\mathbf{W}_h \in \mathbb{R}^{\bar{C}\times C}$, and $\mathbf{W}_d \in \mathbb{R}^{C\times C}$ are implemented as convolutional layers with a kernel size of 1. For the two sub-networks, the self-attention modules can be represented as $\mathbf{p}^{(4)} = \mathbf{p}^{(3)} + \gamma_p \mathbf{r}_p$ and $\mathbf{s}^{(4)} = \mathbf{s}^{(3)} + \gamma_s \mathbf{r}_s$, where $\mathbf{r}_p$ and $\mathbf{r}_s$ are the additional feature maps and $\gamma_p$ and $\gamma_s$ are the corresponding coefficients.

The loss function of the model with two sub-networks is given by $\mathcal{L} = \mathcal{L}_{\mathrm{output}} + \mathcal{L}_{\mathrm{on}}$, where $\mathcal{L}_{\mathrm{output}}$ is the mean squared error (MSE) measuring the overall disaggregation error of the model, and $\mathcal{L}_{\mathrm{on}}$ is the binary cross-entropy (BCE) measuring the classification error of the on/off state classification sub-network. 

\subsection{Training Techniques that Improve Accuracy}

\subsubsection{Training with adversarial loss} 
\begin{figure}[!tb]
\centering
\includegraphics[width=5.5cm]{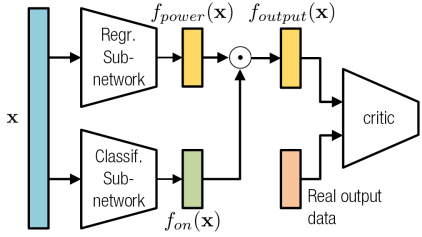}
\caption{The complete SCANet model with the additional critic module when adversarial loss is added.}
\label{framework}
\end{figure}

The performance of SCANet can be further improved by adding an adversarial loss to the model. As is illustrated in Fig. \ref{framework}, a critic network is added to the model so that we can train the model partially as a Wasserstein GAN with gradient penalty (WGAN-GP) \cite{gulrajani2017improved}. The original GAN is formulated by the minimax game between the generator $G$ and the discriminator $D$ \cite{goodfellow2014generative}: 
\begin{equation}
\min_{G}\max_{D} \underset{\mathbf{a}\sim\mathbb{P}_r}{\mathbb{E}}\left[\log\left( D(\mathbf{a}) \right)\right]
+ \underset{\mathbf{\tilde{a}}\sim\mathbb{P}_g}{\mathbb{E}}\left[\log\left( 1 - D(\tilde{\mathbf{a}}) \right)\right],
\end{equation}
where $\mathbb{P}_r$ is the distribution of the real data and $\mathbb{P}_g$ is the distribution of data generated by $\tilde{\mathbf{a}}=G(\mathbf{b}), \mathbf{b}\sim p(\mathbf{b})$, indicating that the input of $G$ is sampled from some noise distribution. Briefly speaking, the goal of the discriminator is to gain the ability to distinguish between real and generated samples, while the generator tries to confuse the discriminator by learning to generate realistic data samples. Training the generator and the discriminator in turn allows the generator gradually obtain the ability to generate realistic samples.

The WGAN-GP model used in this work is modification to the WGAN model proposed in \cite{arjovsky2017wasserstein}, which adopts the Wasserstein distance to stabilize the training of GANs. The gradient penalty in WGAN-GP further stabilizes the training process by penalizing the norm of the gradient of $D$ instead of clipping the weights in $D$ (Here, $D$ is named \emph{critic} instead of \emph{discriminator} as the task of $D$ is not classification of real or generated data). The loss of WGAN-GP (also referred to as the adversarial loss in this paper) is formulated as
\begin{equation}
\begin{split}
\mathcal{L}_{\mathrm{adv}} & = \underset{\tilde{\mathbf{a}}\sim\mathbb{P}_g}{\mathbb{E}}\left[ D\left( \tilde{\mathbf{a}} \right) \right] - \underset{\mathbf{a}\sim\mathbb{P}_r}{\mathbb{E}}\left[ D\left( \mathbf{a} \right) \right] 
\\ & + \lambda \underset{\hat{\mathbf{a}}\sim\mathbb{P}_{\hat{\mathbf{a}}}}{\mathbb{E}} \left[ \left( \left\| \nabla_{\hat{\mathbf{a}}} D \left( \hat{\mathbf{a}} \right) \right\|_2 - 1 \right)^2 \right],
\end{split}
\end{equation}
where the first two terms measure the Wasserstein distance between $\mathbb{P}_r$ and $\mathbb{P}_g$. The last term is the gradient penalty and $\hat{\mathbf{a}}\sim\mathbb{P}_{\hat{\mathbf{a}}}$ refers to uniformly sampling from the line segment connecting point pairs sampled from $\mathbb{P}_r$ and $\mathbb{P}_g$ (see \cite{gulrajani2017improved}). In this paper, instead of generating samples from a noise distribution, we directly use the network producing $f_{\mathrm{output}}(\mathbf{x}_t)$ as the generator. Specifically, we add the adversarial loss $\mathcal{L}_{\mathrm{adv}}$ so that the overall loss function becomes 
\begin{equation}
\mathcal{L} = \mathcal{L}_{\mathrm{output}} + \mathcal{L}_{\mathrm{on}} + \lambda_{\mathrm{adv}} \mathcal{L}_{\mathrm{adv}},
\end{equation}
where $\lambda_{\mathrm{adv}}$ is the weight for the adversarial loss. It is expected that the adversarial loss can help the model produce more realistic output sequences, especially when the size of the training dataset is relatively small.

\begin{table*}
\centering
\captionsetup{justification=centering} 
\caption{Experiment Results on the Evaluation Metrics for the REDD Dataset}
\begin{tabular}{llrrrr} 
\toprule
Metric  & Model & Fridge & Microwave & Dishwasher & Average \\
\midrule
      & Seq2point \cite{zhang2018sequence}  & 26.01      & 27.13      & 24.44      & 25.86       \\
MAE   & SGN \cite{shin2018subtask}          & 26.11      & 16.95      & 15.77      & 19.61       \\
      & Proposed SCANet                     & \bf{21.77} & \bf{13.75} & \bf{10.14} & \bf{15.22}  \\
\midrule
      & Seq2point                           & 16.24      & 18.89      & 22.87      & 19.33       \\
SAE   & SGN                                 & 17.28      & 12.49      & 15.22      & 15.00       \\
      & Proposed SCANet                     & \bf{14.05} & \bf{9.97}  & \bf{8.12}  & \bf{10.71}  \\
\bottomrule
\end{tabular}
\label{REDD}
\end{table*}

\begin{table*}
\centering
\captionsetup{justification=centering} 
\caption{Experiment Results on the Evaluation Metrics for the UK-DALE Dataset}
\begin{tabular}{llrrrrrr}  
\toprule
Metric  & Model          & \tabincell{c}{Washing\\machine}& Kettle & Fridge     & Microwave & Dishwasher   & Average    \\
\midrule
      & Seq2point        & 10.87        & 10.81     & 17.48      & 12.47     & 15.96        & 13.52      \\
MAE   & SGN              & 9.74         & 8.09      & 16.27      & 5.62      & 10.91        & 10.13      \\
      & Proposed SCANet  & \bf{8.48}    & \bf{6.14} & \bf{15.16} & \bf{4.82} & \bf{8.71}    & \bf{8.67}  \\
\midrule
      & Seq2point        & 8.69         & 5.30      & 8.01       & 10.33     & 10.65        & 8.60       \\
SAE   & SGN              & 7.14         & 5.03      & 6.61       & 4.32      & 7.86         & 6.20       \\
      & Proposed SCANet  & \bf{5.77}    & \bf{4.03} & \bf{6.54}  & \bf{3.81} & \bf{4.86}    & \bf{5.00}  \\
\bottomrule
\end{tabular}
\label{UKDALE}
\end{table*}

\subsubsection{On-state augmentation} We propose on-state augmentation to deal with the variance of on-state power consumption of appliances (e.g., the peak power of two fridge models may be different even if they have the same operation pattern). Given an appliance, the maximum offset values $e^- \in \mathbb{R}_-$ and $e^+ \in \mathbb{R}_+$ are decided, and each output sequence $\mathbf{y}$ is replaced by $\mathbf{y} + e \mathbf{o}$, where $e \sim U(e^-, e^+)$. The same amount of on-state offset is also added to the corresponding input sequence $\mathbf{x}$. The augmentation is applied during the training of the model. 

In this work, we apply on-state augmentation to fridge, for which biased estimation of on-state power is a major source of disaggregation error. As expected, the model is able to estimate the on-state power of fridge more accurately after on-state augmentation is implemented.

\section{Results and Analysis}

In this section, we introduce the datasets used in this paper and the implementation details of the models. Experiment results are presented together with empirical analyses of the advantages of SCANet.

\subsection{Experiment Settings}

\subsubsection{Datasets}

Two real-world datasets, REDD \cite{kolter2011redd} and UK-DALE\footnote{http://jack-kelly.com/data/} \cite{kelly2015uk} are used to evaluate the performance of SCANet in this paper. The REDD dataset contains measurement data from six households in the US and the time span of the dataset ranges from 23 to 48 days for different houses. The mains consumption was recorded every 1 second, while the appliance-wise consumption was recorded every 3 seconds. The UK-DALE dataset includes data from five UK households, and measurements for the aggregate consumption as well as consumptions of individual appliances were recorded every 6 seconds. The monitoring of house 1 lasted for over 600 days, while the time spans for the other houses range from 39 to 234 days. Detailed descriptions of the households and the monitored individual appliances in the datasets can be found in \cite{kolter2011redd, kelly2015uk}.

Following previous studies \cite{zhang2018sequence,shin2018subtask}, we use data of houses 2-6 to create the training set and leave the data for house 1 as the test set for the REDD dataset. Disaggregation is implemented for fridge, dishwasher, and microwave. The pre-processed data for the REDD dataset used in this work is provided by the authors of \cite{shin2018subtask}. For the UK-DALE dataset, we use houses 1 and 5 for training, and house 2 for testing. Disaggregation results for fridge, dishwasher, microwave, washing machine, and kettle are reported. In order to normalize the data, we follow the practice in \cite{shin2018subtask} and divide the power consumption values of both datasets by 612, which is the standard deviation of the aggregate consumption for houses 2-5 in the REDD dataset.

\subsubsection{Implementation Details}

\begin{figure*}[!ht]
\centering
\subfloat[Fridge]{
\begin{minipage}[t]{0.33\linewidth}
\centering
\includegraphics[width=5.5cm]{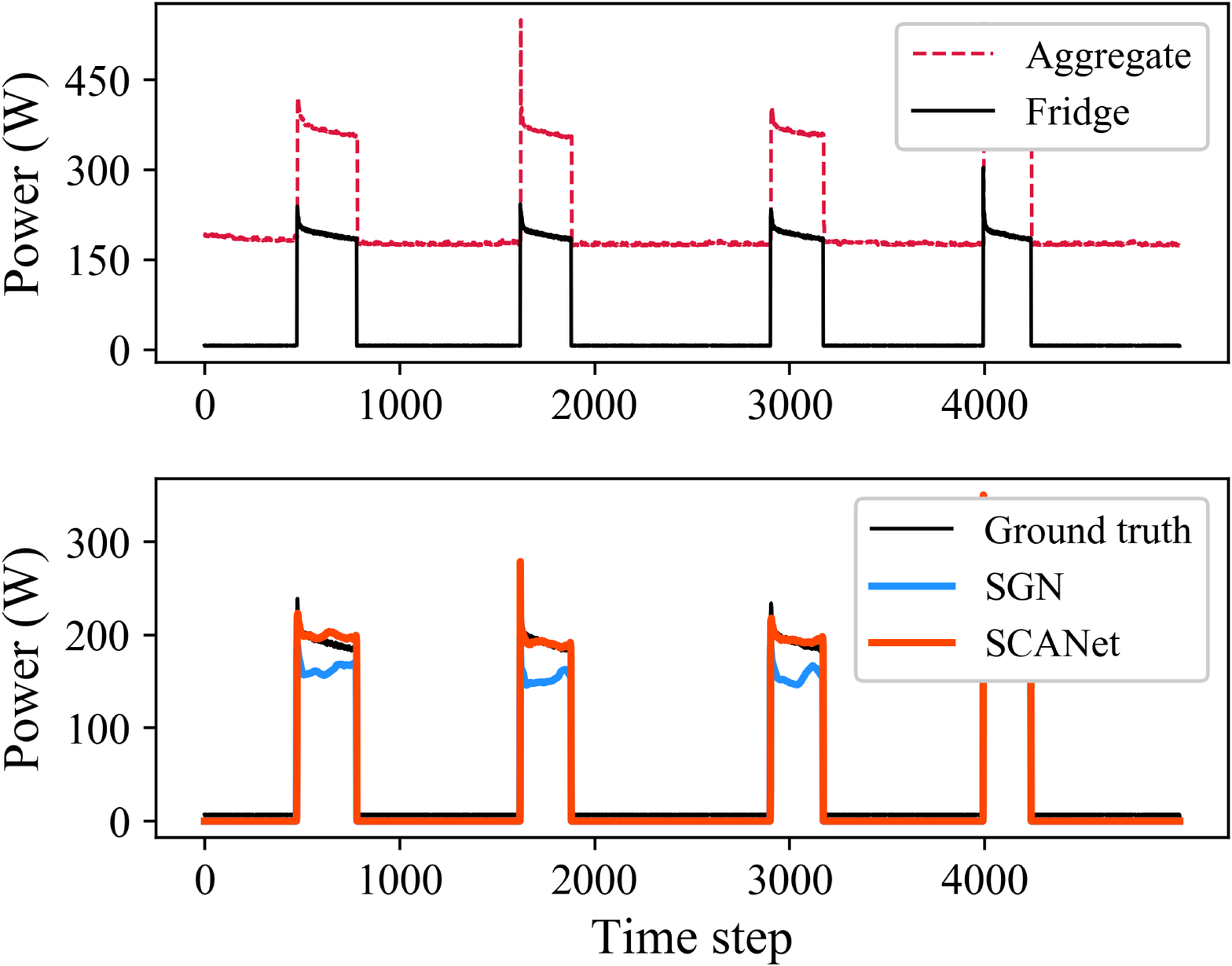}
\end{minipage}%
}%
\subfloat[Microwave]{
\begin{minipage}[t]{0.33\linewidth}
\centering
\includegraphics[width=5.5cm]{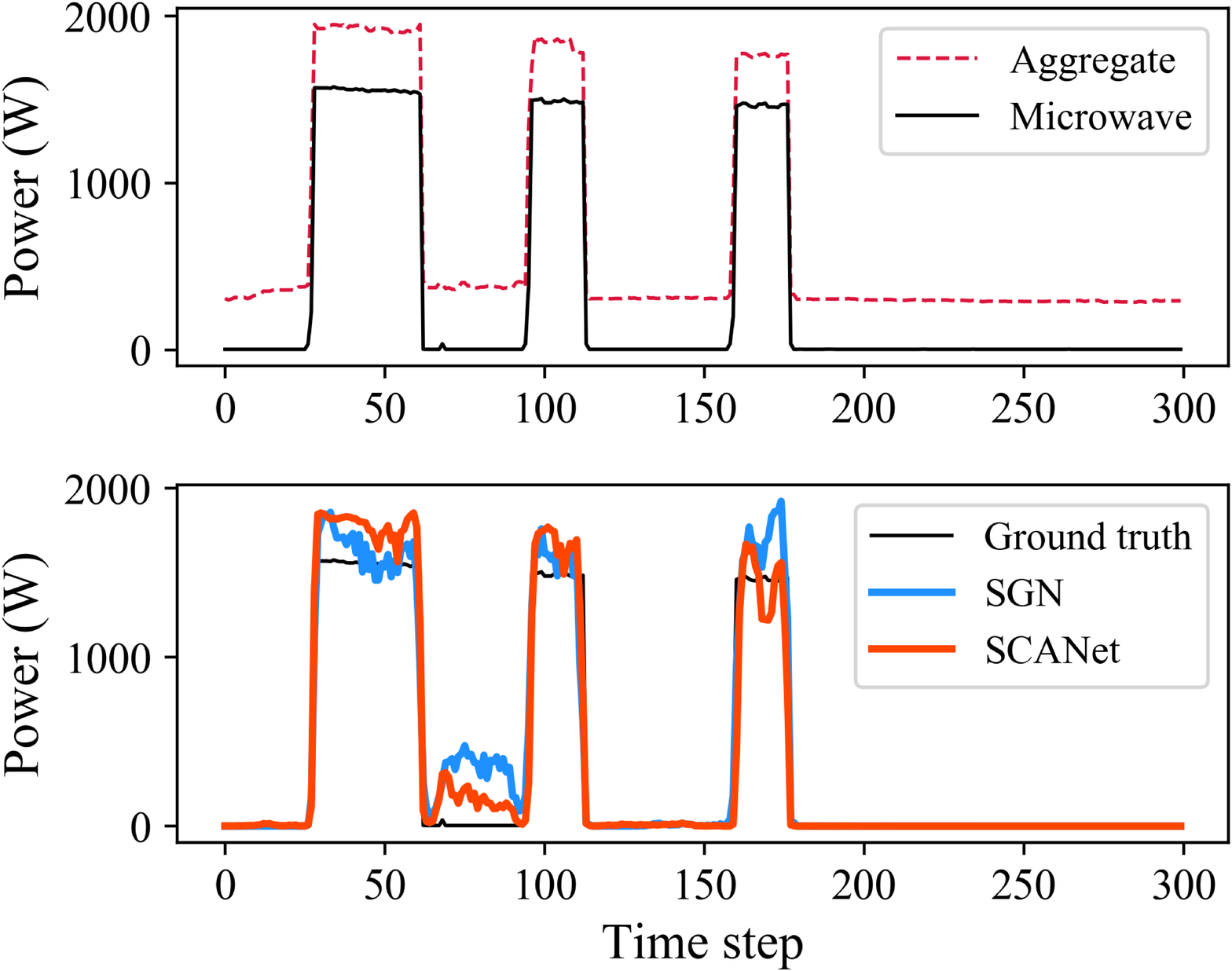}
\end{minipage}%
}%
\subfloat[Dishwasher]{
\begin{minipage}[t]{0.33\linewidth}
\centering
\includegraphics[width=5.5cm]{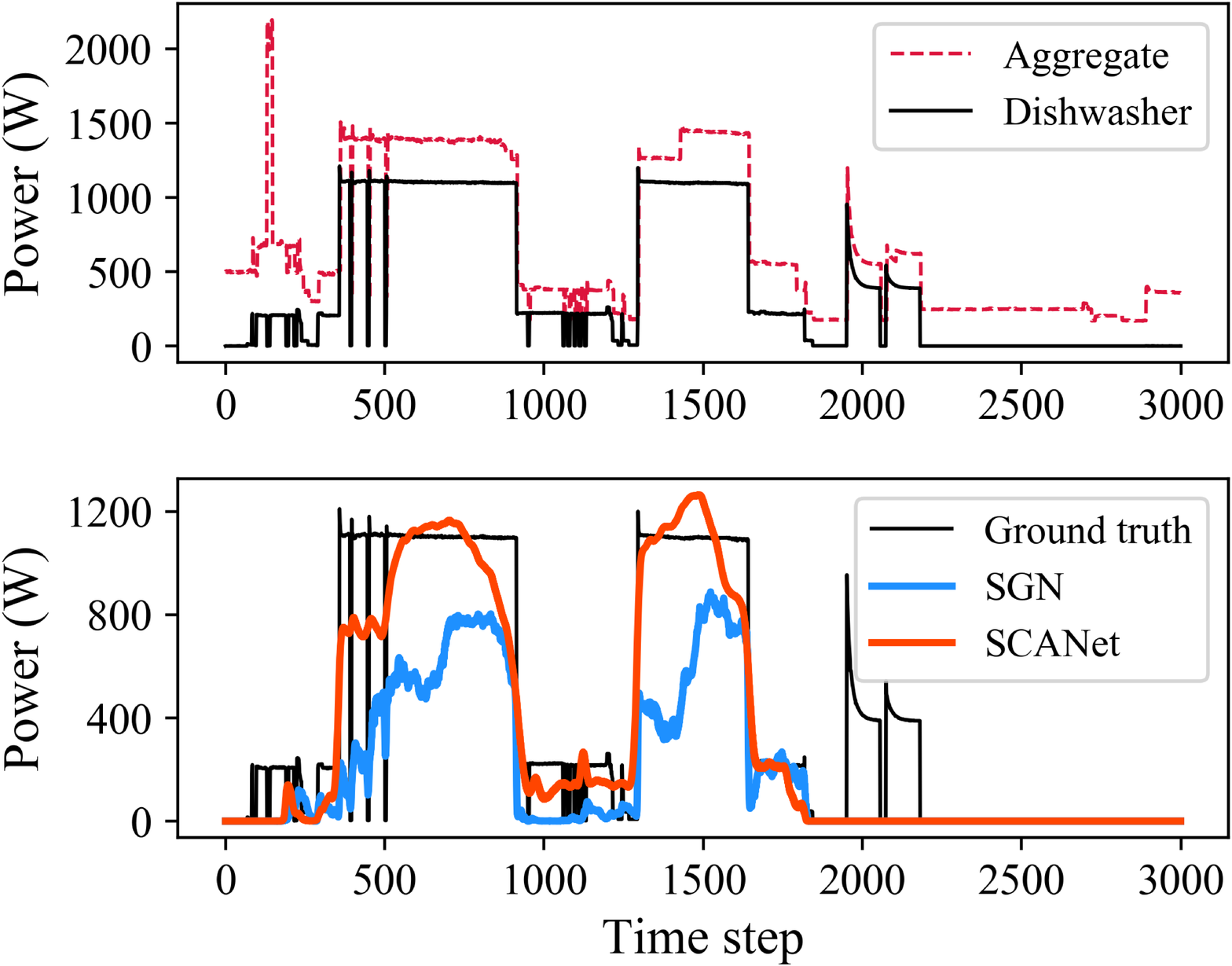}
\end{minipage}%
}%
\caption{Samples of disaggregation results for the REDD dataset.}
\label{result_REDD}
\end{figure*}

\begin{figure*}[!ht]
\centering
\subfloat[Fridge]{
\begin{minipage}[t]{0.33\linewidth}
\centering
\includegraphics[width=5.5cm]{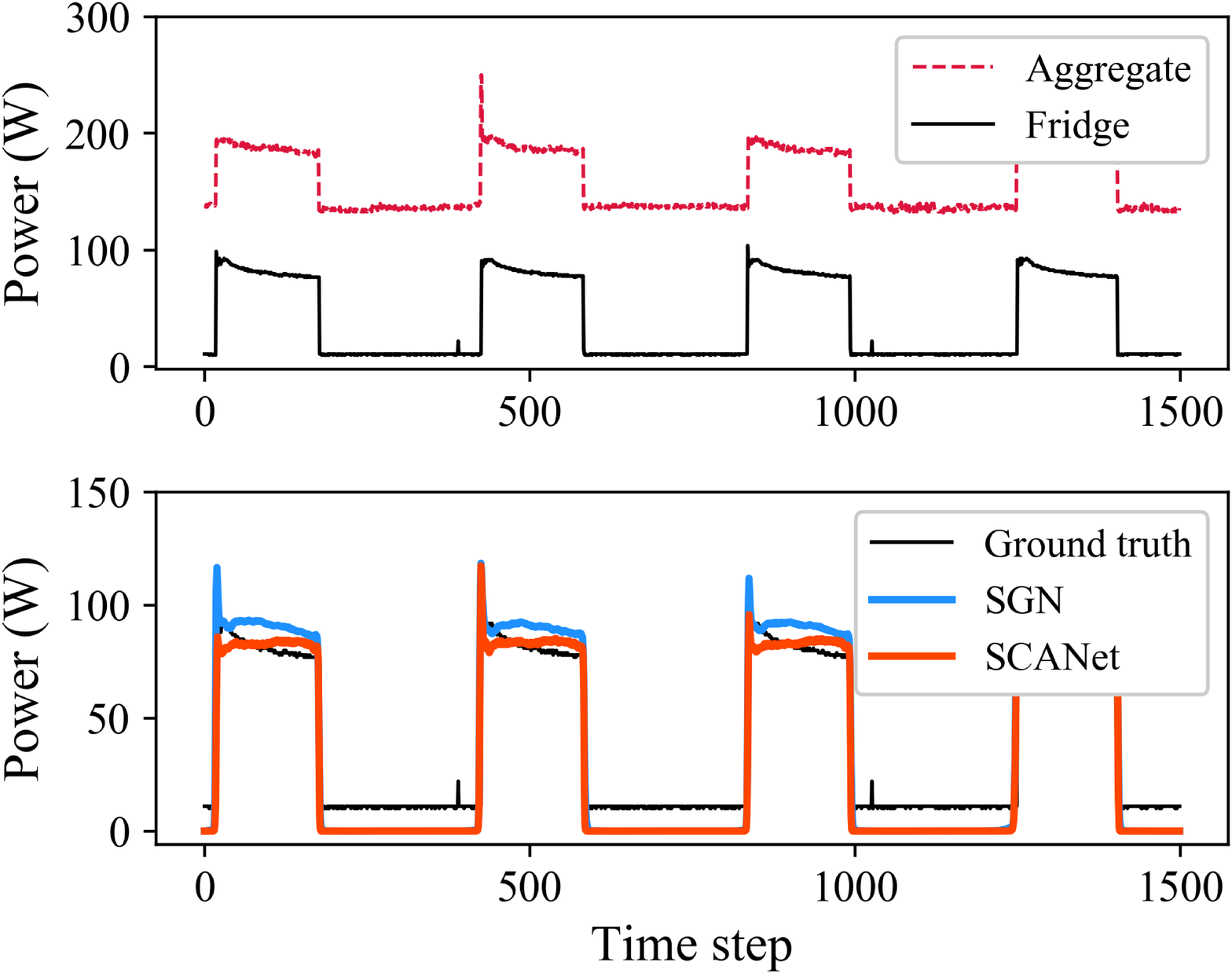}
\end{minipage}%
}%
\subfloat[Microwave]{
\begin{minipage}[t]{0.33\linewidth}
\centering
\includegraphics[width=5.5cm]{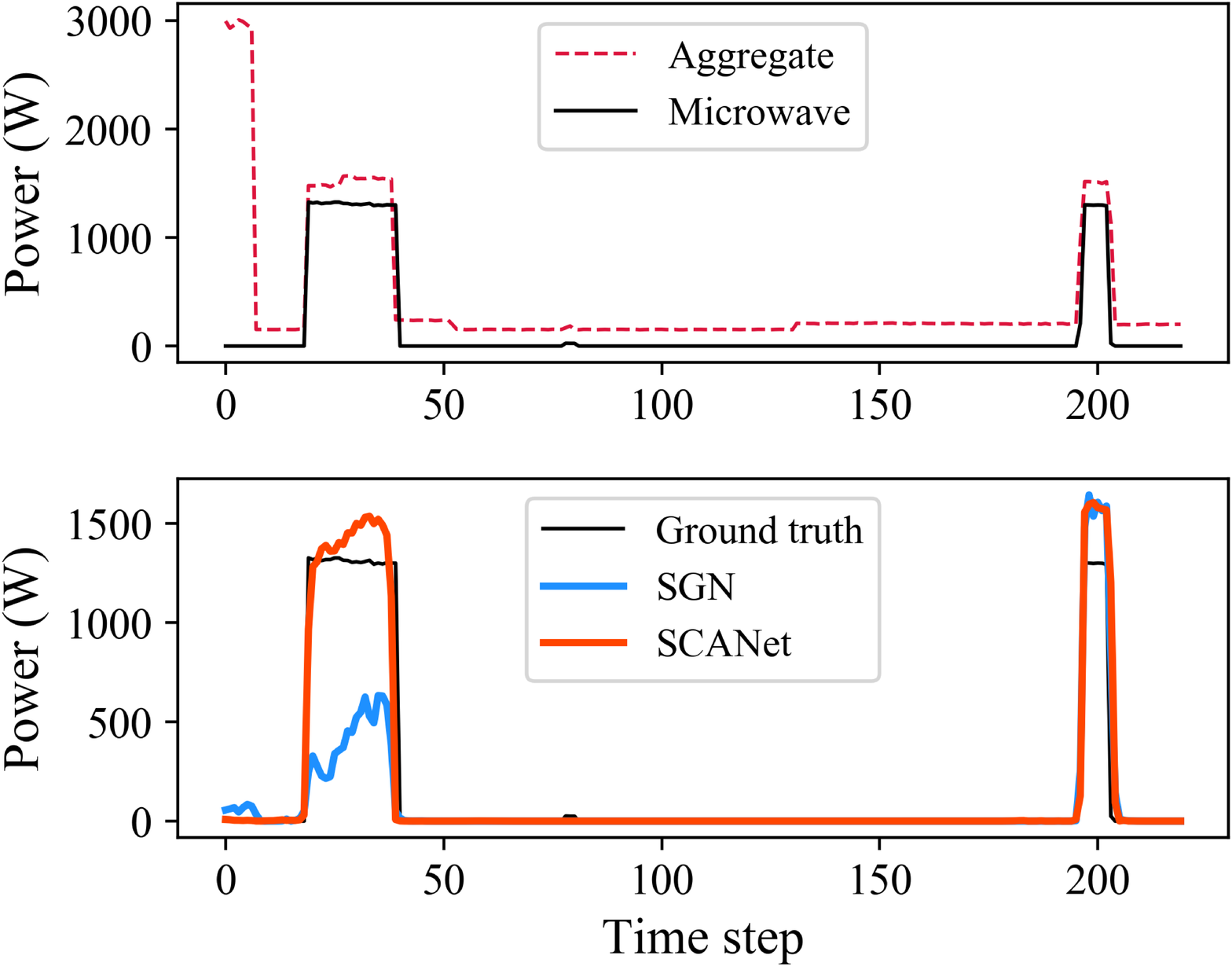}
\end{minipage}%
}%
\subfloat[Dishwasher]{
\begin{minipage}[t]{0.33\linewidth}
\centering
\includegraphics[width=5.5cm]{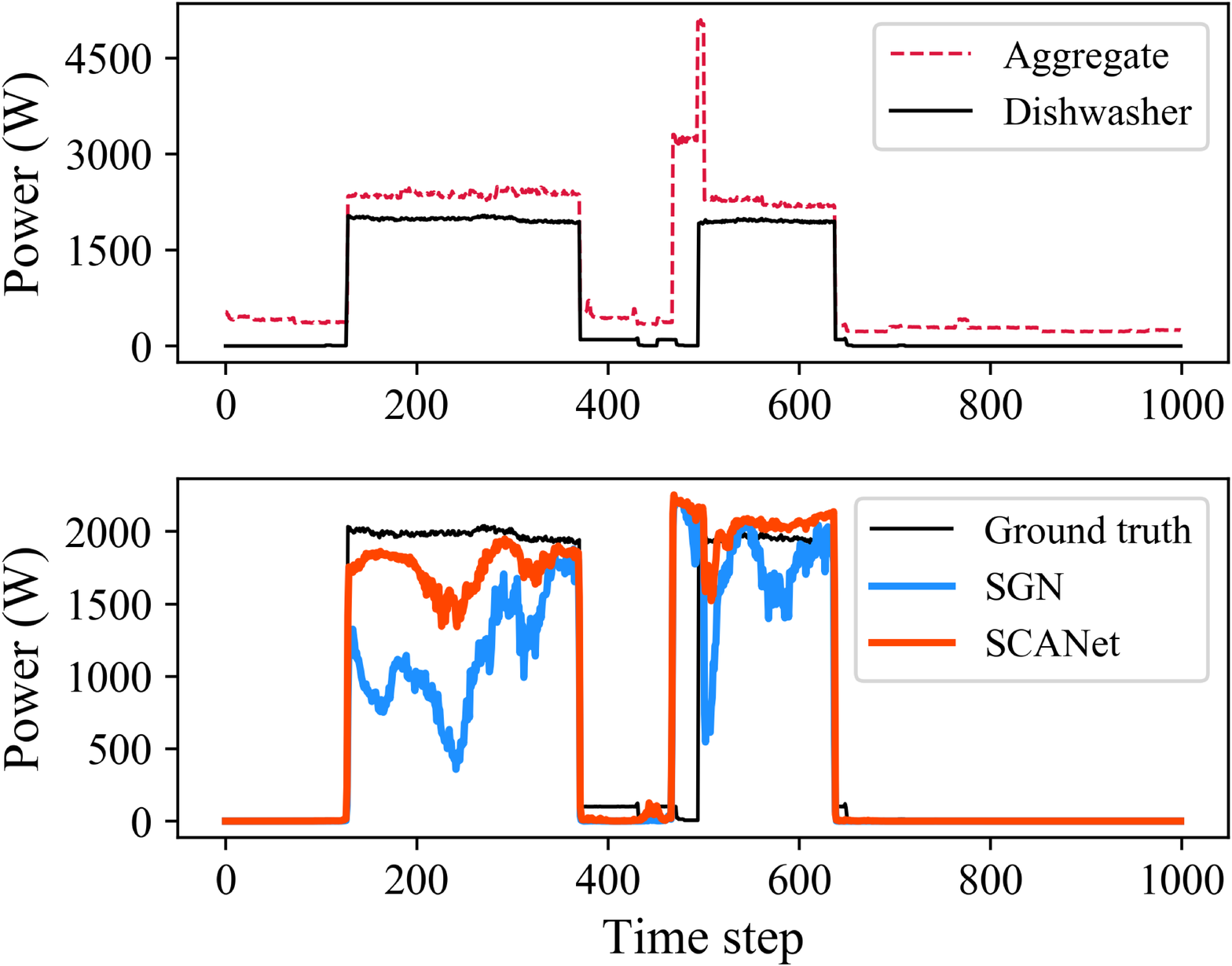}
\end{minipage}%
}%
\caption{Samples of disaggregation results for the UK-DALE dataset.}
\label{result_UK}
\end{figure*}

\begin{figure}[tbh]
\centering
  \subfloat[A false positive example of SGN]{
    \includegraphics[width=.72\columnwidth]{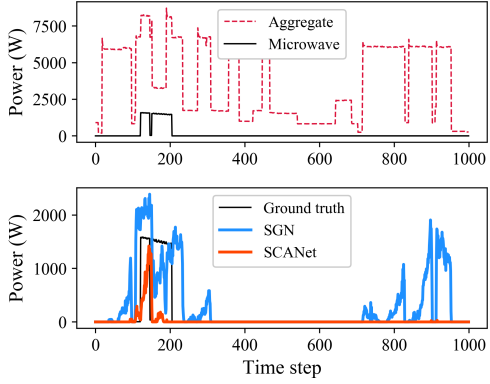}
  }\\
  \subfloat[A case our model is able to detect]{
    \includegraphics[width=.72\columnwidth]{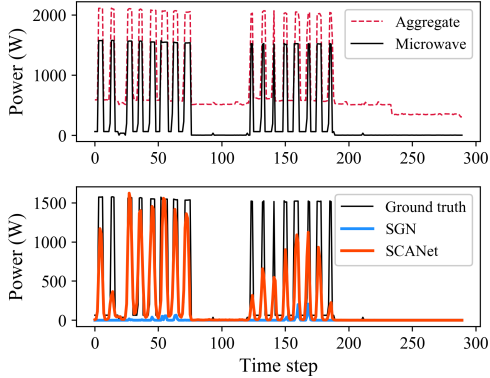}
  }  
\caption{Additional samples of disaggregation results for microwave in the REDD dataset.}
\label{microwave_REDD}
\end{figure}

\begin{figure}[!tb]
\centering
\includegraphics[width=8.5cm]{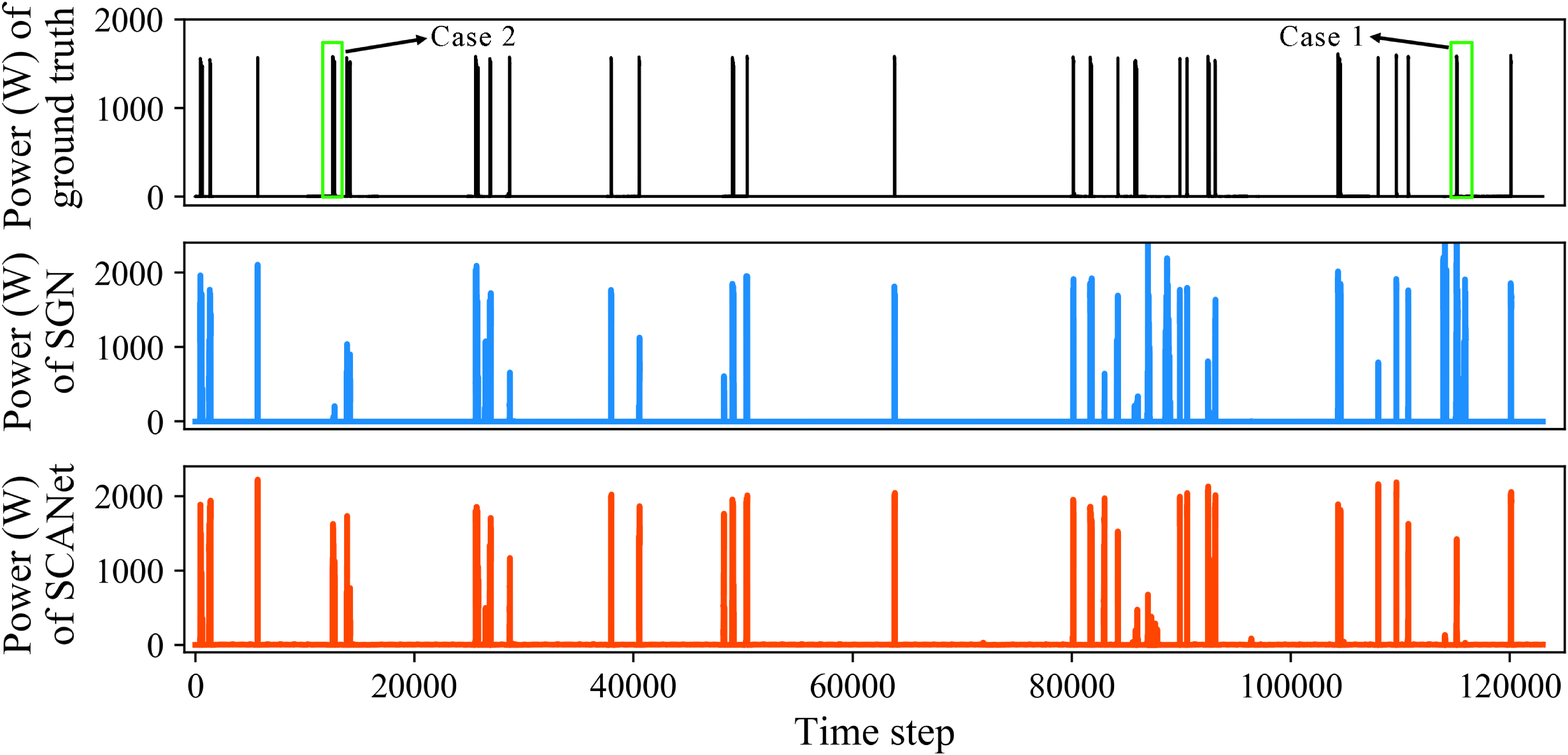}
\caption{Comparison of the performance of SGN and SCANet for microwave in the REDD dataset. The time span for the figure is roughly 100 hours.}
\label{microwave_compare}
\end{figure}

The SGN backbone \cite{shin2018subtask} has 6 convolutional layers followed by 2 FC layers in each sub-network. Specifically, the numbers of filters in each layer are 30, 30, 40, 50, 50, and 50, and the kernel sizes are 10, 8, 6, 5, 5, and 5, respectively. All convolutional layers are implemented with a stride of 1 and ``same'' padding, and the weights are initialized with ``He normal'' initializer \cite{he2015delving}. The first FC layer has 1024 hidden nodes, and the second FC layer has the same number of nodes as the output of the model. All but the last layer uses ReLU activation function. For the REDD dataset, the output sequence size $s$ is 64 and the additional window size $w$ is 400, while $s$ and $w$ are 32 and 200 for the UK-DALE dataset. As the sizes of the input and the output are reduced by half for the UK-DALE dataset, we also change the kernel sizes to 5, 4, 3, 3, 3, and 3 while other hyper-parameters remain the same. Adam optimizer with initial learning rate 0.0001 is adopted and we train the models for 5 epochs with a batch size of 16. For SCANet, we add two parallel branches with $r_d=2$ and $r_d=3$ to each sub-network starting from the 4th convolutional layer. The layer producing $\mathbf{p}^{(3)}$ and $\mathbf{s}^{(3)}$ has 64 filters, thus $C=64$ for the self-attention modules, and we set $\bar{C}$ to 32. We adopt most of the hyper-parameters from \cite{shin2018subtask} such that we can focus on ensuring the effectiveness of the proposed model components rather than tuning the hyper-parameters. All models are implemented in Python 3.6 with Keras 2.1.6.

The input samples are produced by a sliding window running over the input sequences with specific step sizes, which is set to 2 for the REDD dataset. The step sizes for microwave, dishwasher, fridge, washing machine, and kettle are 4, 8, 32, 32, and 32 for the UK-DALE dataset. We choose the step sizes by ensuring that the SGN model performs no worse than its reported results \cite{shin2018subtask}. For testing, a sliding window with a step size of 2 generates the input samples. Multiple overlapping output sequences are averaged to produce the final output. Further, as on-state events are relatively rare for some appliances, the imbalance of on and off states may bring difficulties to the training of the models. Thus, we randomly remove off-state samples from the training dataset for some appliances. For the REDD dataset, the probablity of keeping an off-state sample (i.e., the entire output of the sample is off-state) is 0.2 for dishwasher. The probabilities for dishwasher, microwave, and kettle are 0.02, 0.05, and 0.1, respectively, for the UK-DALE dataset. The same settings are shared by the Seq2point model \cite{zhang2018sequence} and SGN when applicable.

For the implementation of WGAN-GP, a simple critic with 4 convolutional layers and 32 filters at each layer is formulated. The kernel sizes are set to 3, an FC layer with 256 hidden nodes bridges the convolutional layers and the output node, and the weight $\lambda_{\mathrm{adv}}$ for $\mathcal{L}_{\mathrm{adv}}$ is 0.5. We train the model with a batch size of 32. Specifically, we implement the model with $\mathcal{L}_{\mathrm{adv}}$ for appliances other than fridges. On-state augmentation is implemented in the training of the model for fridges with $e^-=-0.1$ and $e^+=0.1$ for the REDD dataset and $e^-=-0.03$ and $e^+=0.03$ for the UK-DALE dataset.

Mean absolute error (MAE) and signal aggregate error (SAE) are used as the evaluation metrics for each appliance \cite{shin2018subtask}. Specifically, given a predicted output sequence with $T$ time steps, ${\rm SAE} = \frac{1}{N}\sum^N_{\tau=1}{\frac{1}{M} |r_\tau - \hat{r}_\tau |}$, where $N$ is the number of disjoint time periods with length $M$, $T=N\times M$, $\hat{r}_\tau$ is the total predicted power consumption in the $\tau$th time period, and $r_\tau$ is the corresponding ground truth. In this work, we set $N=1200$, thus each time period corresponds to approximately one hour for the REDD dataset, and two hours for the UK-DALE dataset.

\subsection{Experiment Results}

\begin{figure}[!tb]
\centering
\captionsetup{justification=centering}
\includegraphics[width=8cm]{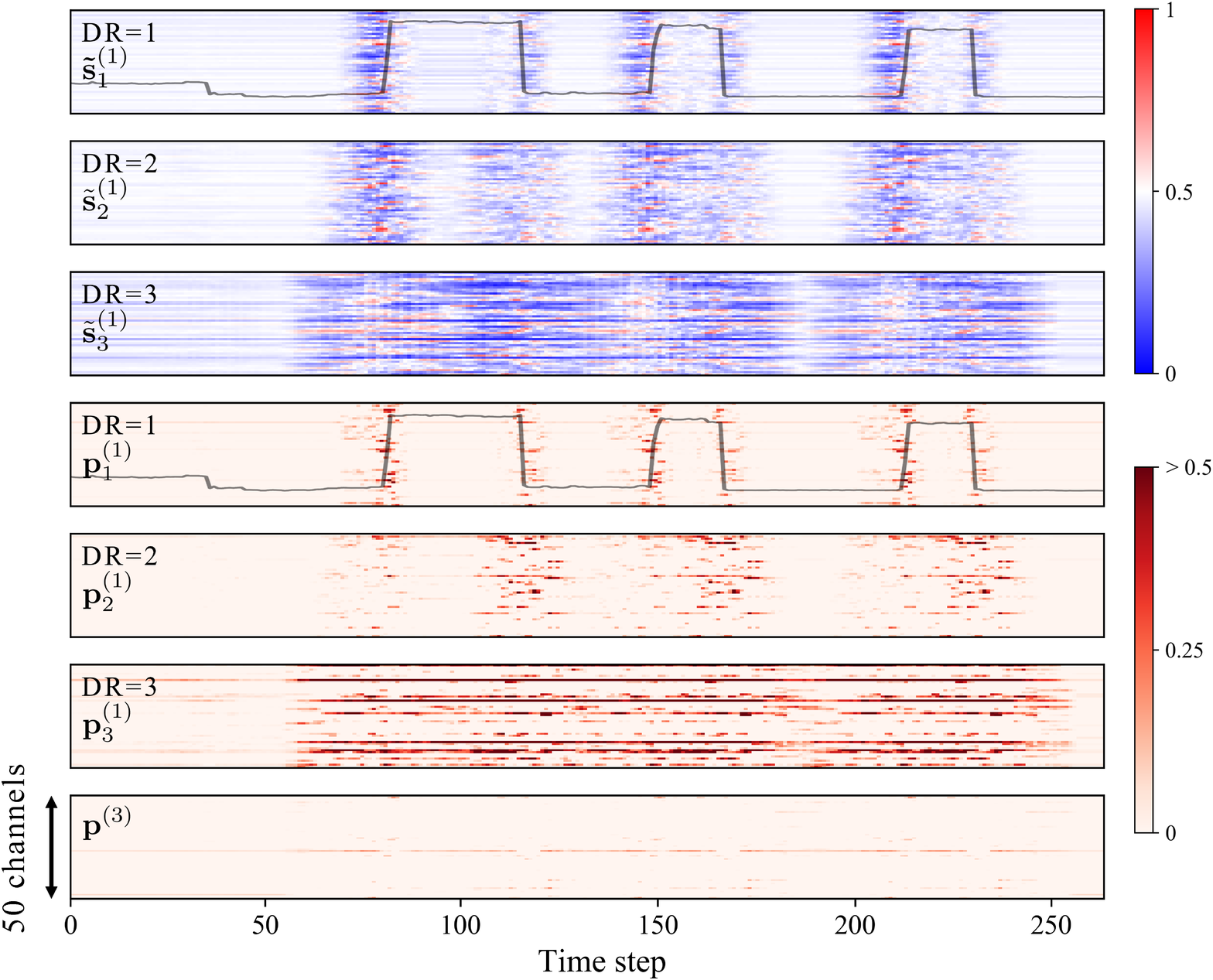}
\caption{Visualization of feature maps of multiple scales for microwave in the REDD dataset. $\tilde{\mathbf{s}}^{(1)}_1$ to $\tilde{\mathbf{s}}^{(1)}_3$ and $\mathbf{p}^{(1)}_1$ to $\mathbf{p}^{(1)}_3$ are the feature maps at the end of the three branches in the classification sub-network and the regression sub-network, respectively. Note that the sigmoid function is used for $\tilde{\mathbf{s}}^{(1)}_1$ to $\tilde{\mathbf{s}}^{(1)}_3$ as the activation function.}
\label{scale}
\end{figure}

\begin{figure}[tbh]
\centering
  \subfloat[The attention matrix for Fig. \ref{result_REDD} (b)]{
    \includegraphics[width=.72\columnwidth]{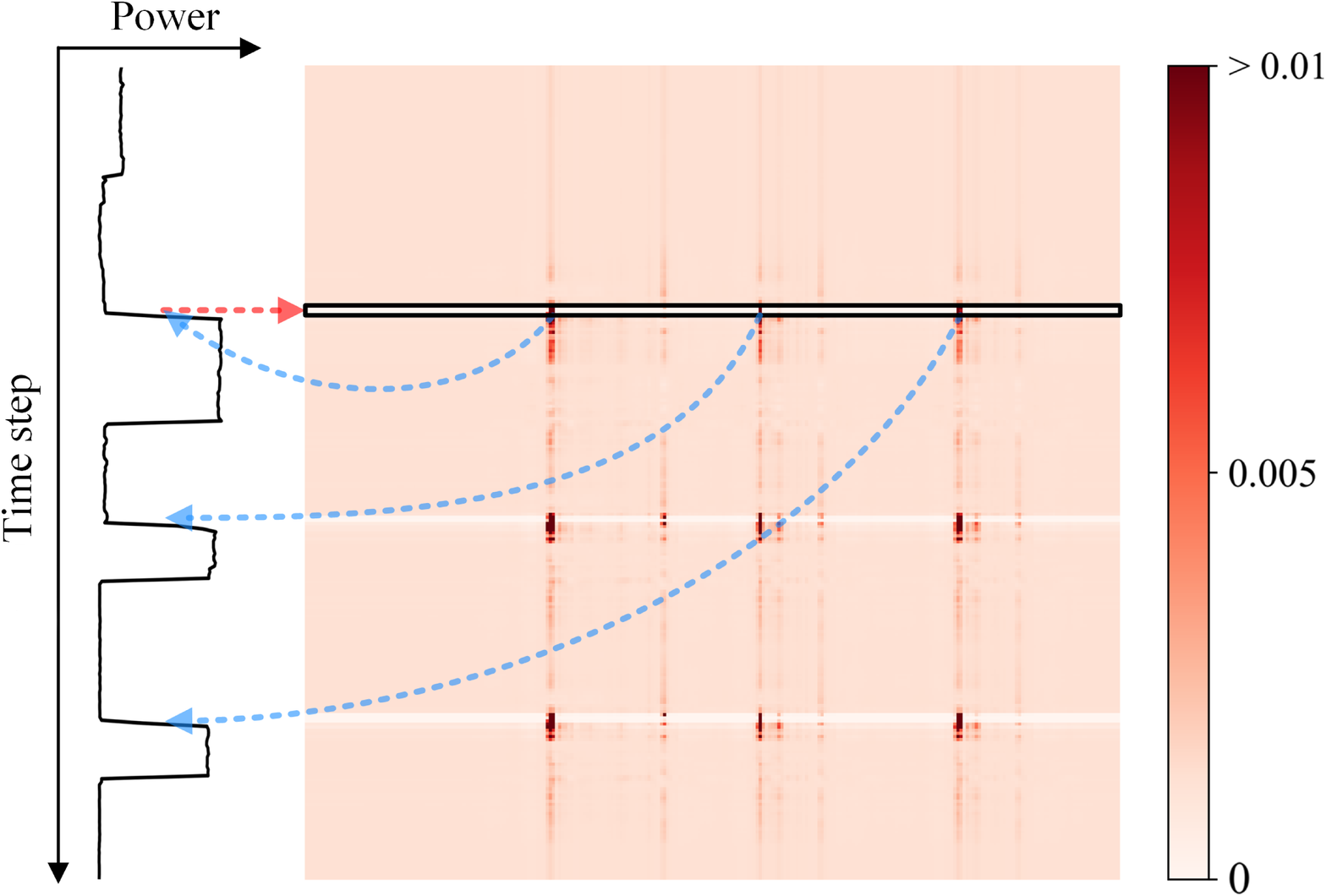}
  }\\
  \subfloat[The attention matrix for Fig. \ref{microwave_REDD} (b)]{
    \includegraphics[width=.72\columnwidth]{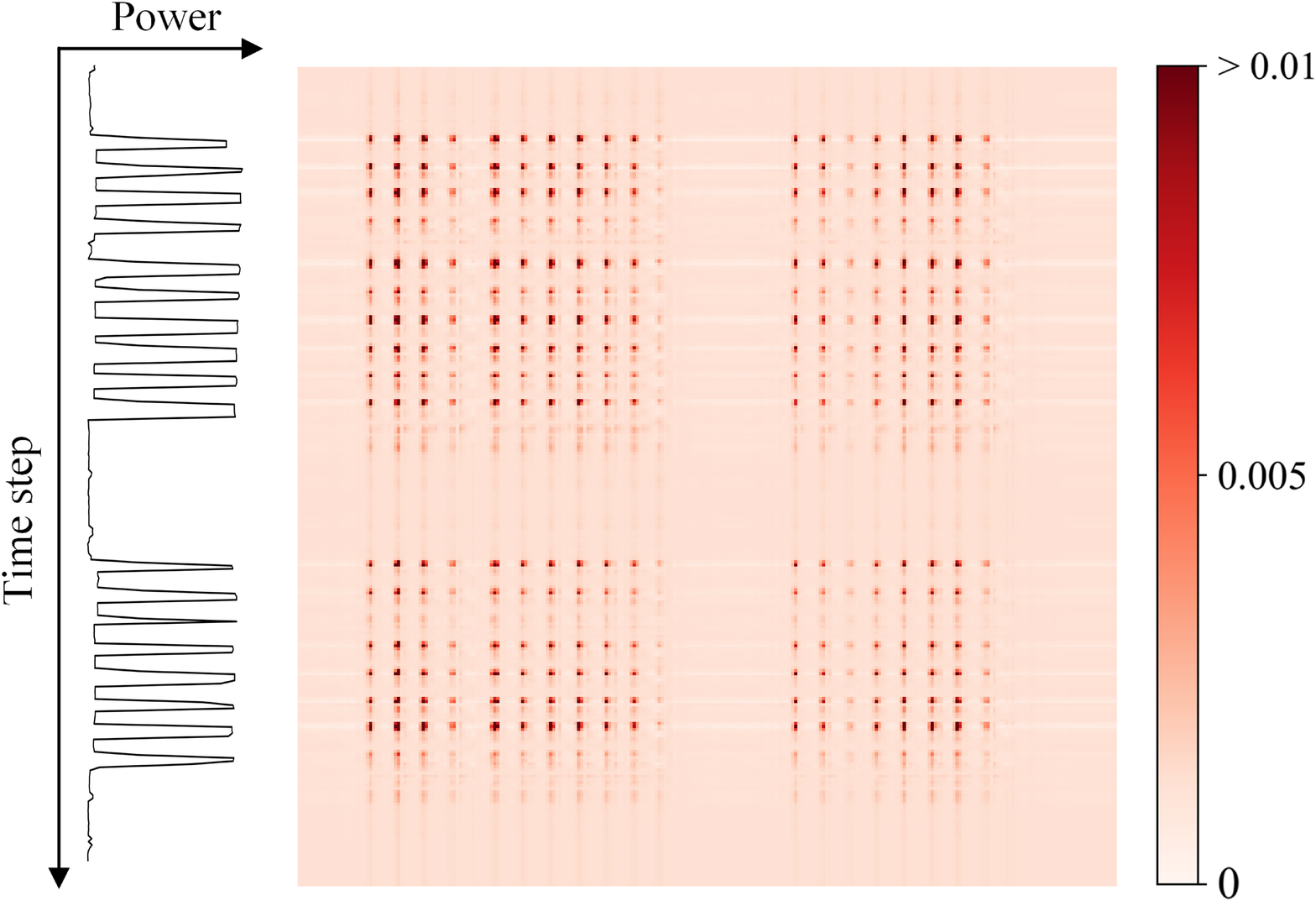}
  }  
\caption{Visualization of the self-attention matrix $\mathbf{A}^\top$ in the on/off state classification sub-network for microwave in the REDD dataset.}
\label{attention}
\end{figure}

\begin{figure}[!tb]
\centering
\includegraphics[width=7.5cm]{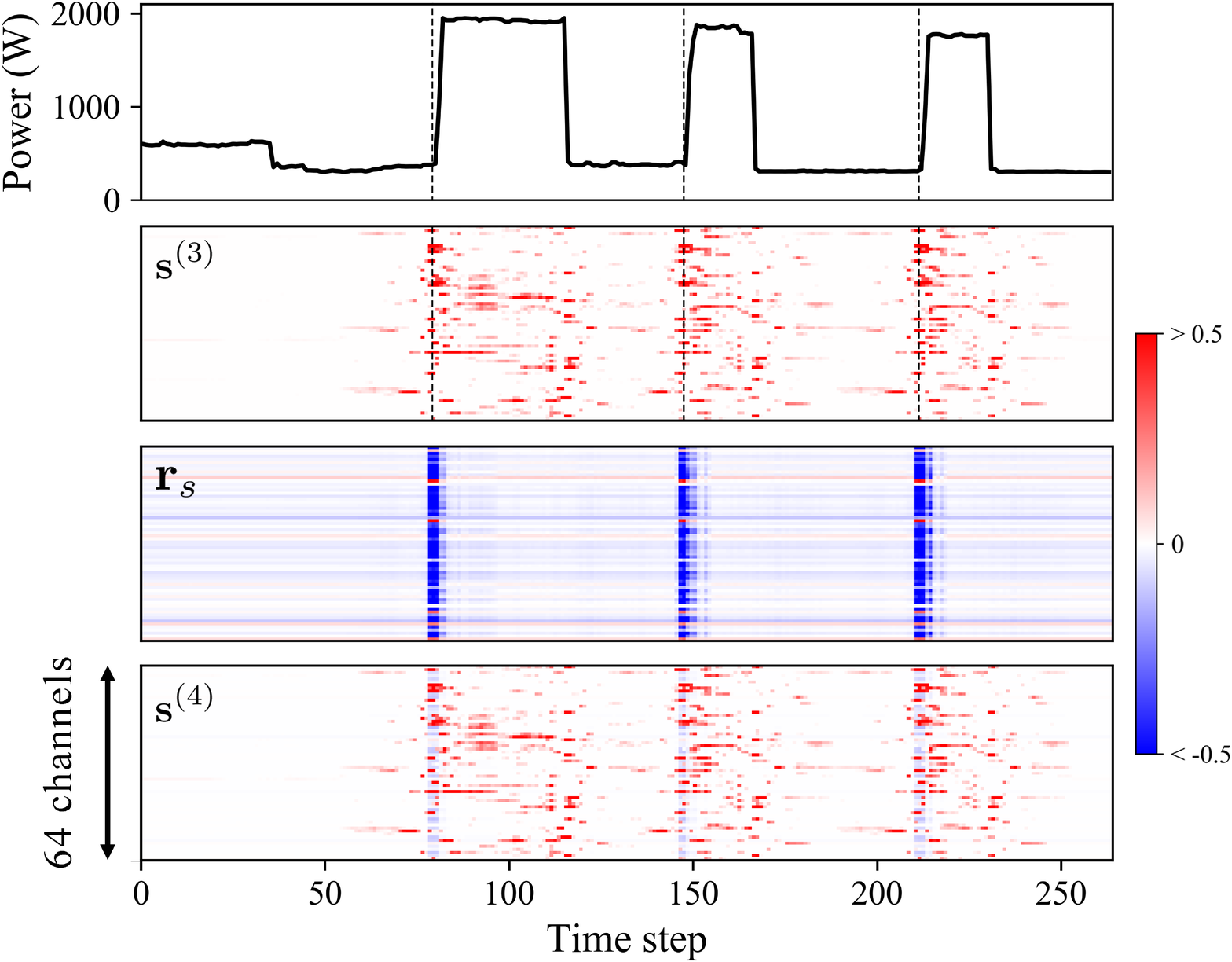}
\caption{Visualization of feature maps in the on/off state classification sub-network for microwave in the REDD dataset.}
\label{edge}
\end{figure}

\begin{figure}[!tb]
\centering
\includegraphics[width=.75\columnwidth]{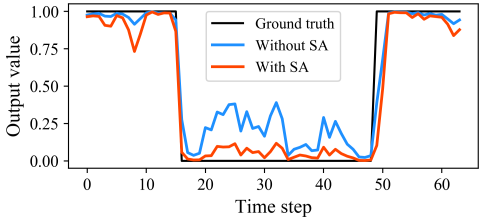}
\caption{Outputs of the on/off state classification sub-network with and without the SA module for the input in Fig. \ref{edge}.}
\label{SA}
\end{figure}

\begin{figure}[!tb]
\centering
\includegraphics[width=7.5cm]{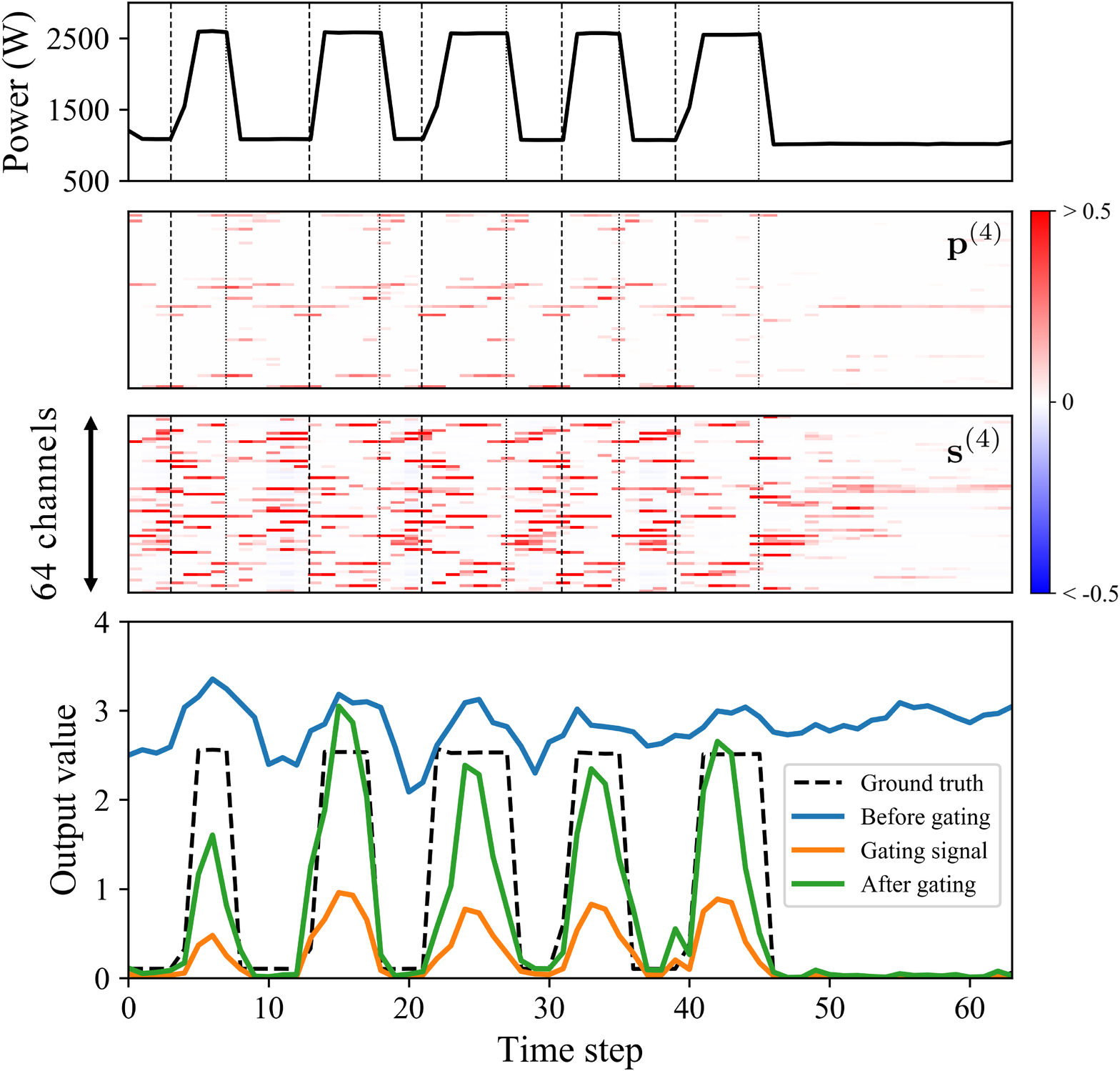}
\caption{Visualization of feature maps of a rare case for microwave in the REDD dataset. Two types of vertical dashed lines are added to highlight the rising and falling edges in the first three sub-figures.}
\label{rare}
\end{figure}

We report the results of the evaluation metrics for the REDD and the UK-DALE datasets in Table \ref{REDD} and Table \ref{UKDALE}, respectively. Each value is obtained by averaging results from 3 trials. It is seen in the two tables that SCANet achieves lower MAEs and SAEs than SGN for all of the appliances, especially for the REDD dataset, for which the improvements are 22.39\% and 28.60\% for averaged MAEs and SAEs. The average improvements for the UK-DALE dataset are also greater than 10\%.

The comparison of SCANet with SGN is further presented in Fig. \ref{result_REDD} and Fig. \ref{result_UK}, and we can see that SCANet produces more accurate disaggregation results than SGN. In Fig. \ref{result_REDD} (a), it is observed that on-state augmentation helps the model capture the on-state power consumption of fridges. We further showcase the advantage of SCANet in Fig. \ref{microwave_REDD}, in which two cases are illustrated for microwave in the REDD dataset. A false positive case of SGN is shown in Fig. \ref{microwave_REDD} (a). Fig. \ref{microwave_REDD} (b) shows that SCANet is able to identify that the microwave consumes power for multiple short durations successively while SGN fails to tell that the microwave is on. In Fig. \ref{microwave_compare}, the disaggregation results of the SGN and the SCANet models for microwave are illustrated. The two cases in Fig. \ref{microwave_REDD} are also marked in Fig. \ref{microwave_compare}. In general, the SCANet model is more accurate in terms of power consumption level and has fewer false false positive cases.

The activations at the end of the branches in the two sub-networks for the case in Fig. \ref{result_REDD} (b) are visualized in Fig. \ref{scale}. For the sample used for the visualization, only the 256 time steps in the middle are plotted. Specifically, each feature map contains values of 256 time steps (the horizontal axis) and 50 channels (the longitudinal axis). It is clear that the branches are all responding to the rising and falling edges in the microwave consumption signal (the signal is added to the feature maps of $\tilde{\mathbf{s}}^{(1)}_1$ and $\mathbf{p}^{(1)}_1$ for comparison), and that a large proportion of the gating signals are actually suppressing the high activations in the regression sub-network ($\tilde{\mathbf{s}}^{(1)}_1$ to $\tilde{\mathbf{s}}^{(1)}_3$ are used as gating signals for $\mathbf{p}^{(1)}_1$ to $\mathbf{p}^{(1)}_3$). As a result, the feature map $\mathbf{p}^{(3)}$ is much less activated in general.

We also use the cases in Fig. \ref{result_REDD} (b) and Fig. \ref{microwave_REDD} (b) to illustrate the mechanism of the self-attention module. We first visualize the attention matrix in the classification sub-network for the case of Fig. \ref{result_REDD} (b) in Fig. \ref{attention} (a). Clearly, the model mainly focuses on the edges in this time range (note that the assignment of attention is row-wise), and the highest attention values are observed for the three rising edges, i.e., the model refers to all the rising edges when looking at one of the rising edges. For instance, we highlight the row corresponding to the first rising edge in $\mathbf{A}^\top$, and the high values in the row mainly belong to the three rising edges in the signal including the first rising edge itself. Further, the attention matrix for Fig. \ref{microwave_REDD} (b) is shown in Fig. \ref{attention} (b). Similarly, the main feature of the matrix is that high attention values are found for the time steps corresponding to rising edges and the rising edges are attending to all rising edges within the time range.

In Fig. \ref{edge}, we plot the input and output of the self-attention module for the case of Fig. \ref{result_REDD} (b) for the classification sub-network as well as the additional feature map $\mathbf{r}_s$, which is highly activated at all three rising edges. As $\mathbf{s}^{(4)} = \mathbf{s}^{(3)} + \gamma_s \mathbf{r}_s$ and $\lambda_s=0.0764$ for the sample, the activations at the edges are reflected in $\mathbf{s}^{(4)}$. Thus, it is of interest to investigate the effect of $\lambda_s \mathbf{r}_s$ in $\mathbf{s}^{(4)}$. To this end, we bypass the self-attention network and directly feed $\mathbf{s}^{(3)}$ to the FC layers to obtain the output of the classification sub-network and compare it with the original output in Fig. \ref{SA}, which shows that $\mathbf{r}_s$ helps suppress the on-state probability where the microwave is not consuming electricity. Note that this can be a hard task as the microwave is turned on shortly before and after. Thus, in this case, the regression sub-network only needs to produce an output sequence with approximately the same value and the classification sub-network alone is able to produce desirable results. 

We use the case in Fig. \ref{microwave_REDD} (b) to show that the regression sub-network is not obsolete. Specifically, we plot the feature maps $\mathbf{p}^{(4)}$ and $\mathbf{s}^{(4)}$, the outputs of the sub-networks as well as the output of the model in Fig. \ref{rare}, which shows that $\mathbf{p}^{(4)}$ is actively responding to both rising and falling edges of the input, forming a repetitive pattern. As a result, the output of the regression sub-network predicts the right trends of the power consumption in the time period.

\subsection{Ablation Study and Discussion}

\begin{table}[!tb]
\centering
\caption{Ablation Study Results for Appliances in the REDD Dataset in Terms of MAE.  MS, SA, AL, and OA Are the Acronym for ``Multi-scale'', ``Self-attention'', ``Adversarial Loss'', and ``On-state Augmentation''}
\begin{tabular}{ccccrrr}  
\toprule

MS         & SA         & AL         & OA			& Dishwasher    & Microwave 	&  Fridge		\\
\midrule
-          & -          & -          & -			& 15.77         & 16.95    		&  26.11 		\\
\checkmark &            &            & 				& 13.22         & 15.93    		&  25.52		\\
           & \checkmark &            & 				& 13.53         & 15.98    		&  25.02		\\
\checkmark & \checkmark &            & 				& 12.73         & 15.00    		&  24.68		\\
           &            & \checkmark & 				& 13.54         & 15.97    		&  -			\\
           &            &            & \checkmark 	& -	        	& -    			&  22.80 		\\           
\checkmark & \checkmark & \checkmark & 				& \bf{10.14}    & \bf{13.75}   	&  -		 	 \\
\checkmark & \checkmark & 			 & \checkmark 	& -			    & - 		  	&  \bf{21.77} 	 \\

\bottomrule
\end{tabular}
\label{ablation}
\end{table}

\begin{figure*}[tbh]
\centering
\subfloat[]{
\begin{minipage}[t]{0.3\linewidth}
\centering
\includegraphics[width=5cm]{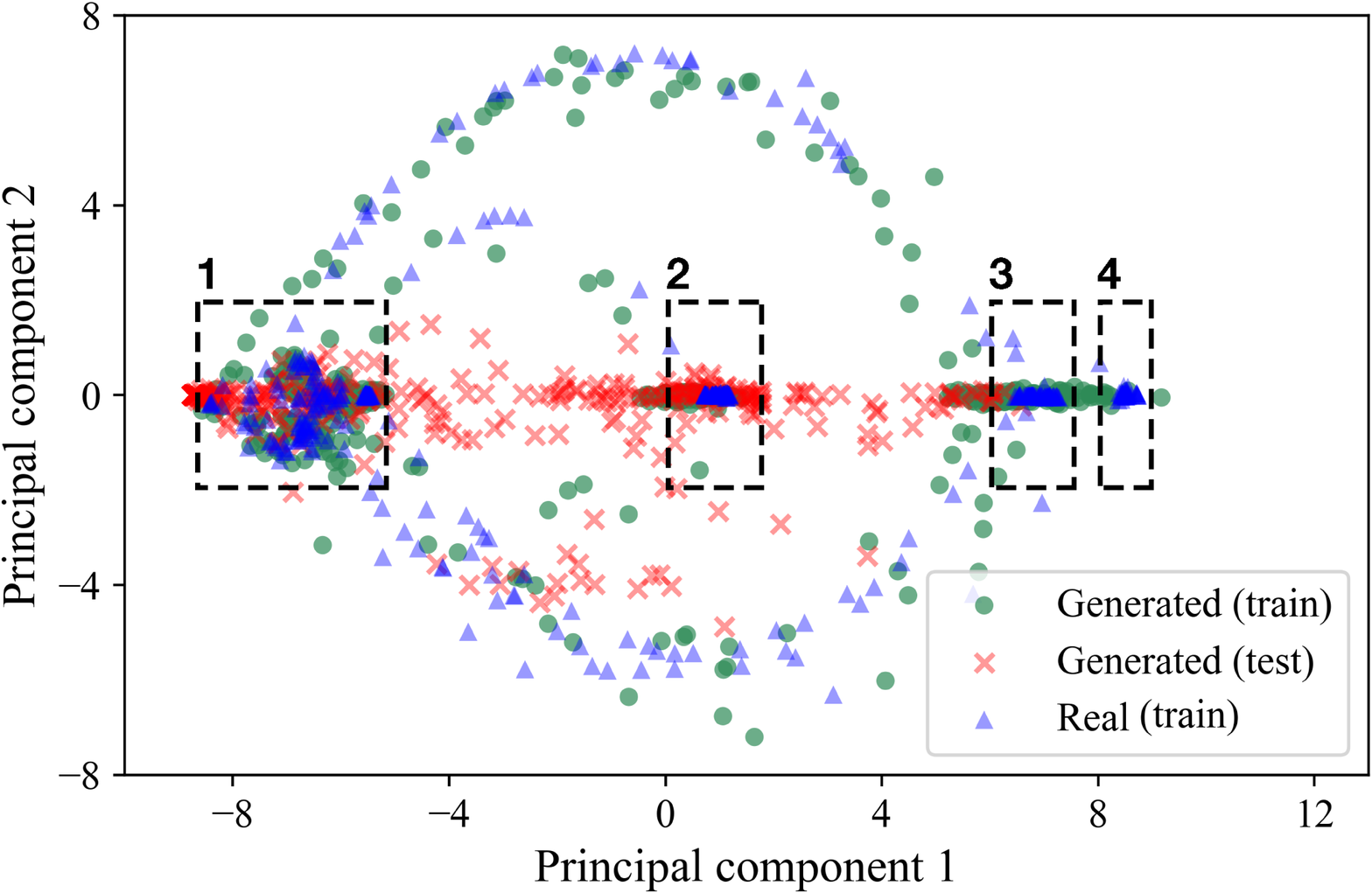}
\end{minipage}%
}%
\subfloat[]{
\begin{minipage}[t]{0.3\linewidth}
\centering
\includegraphics[width=5cm]{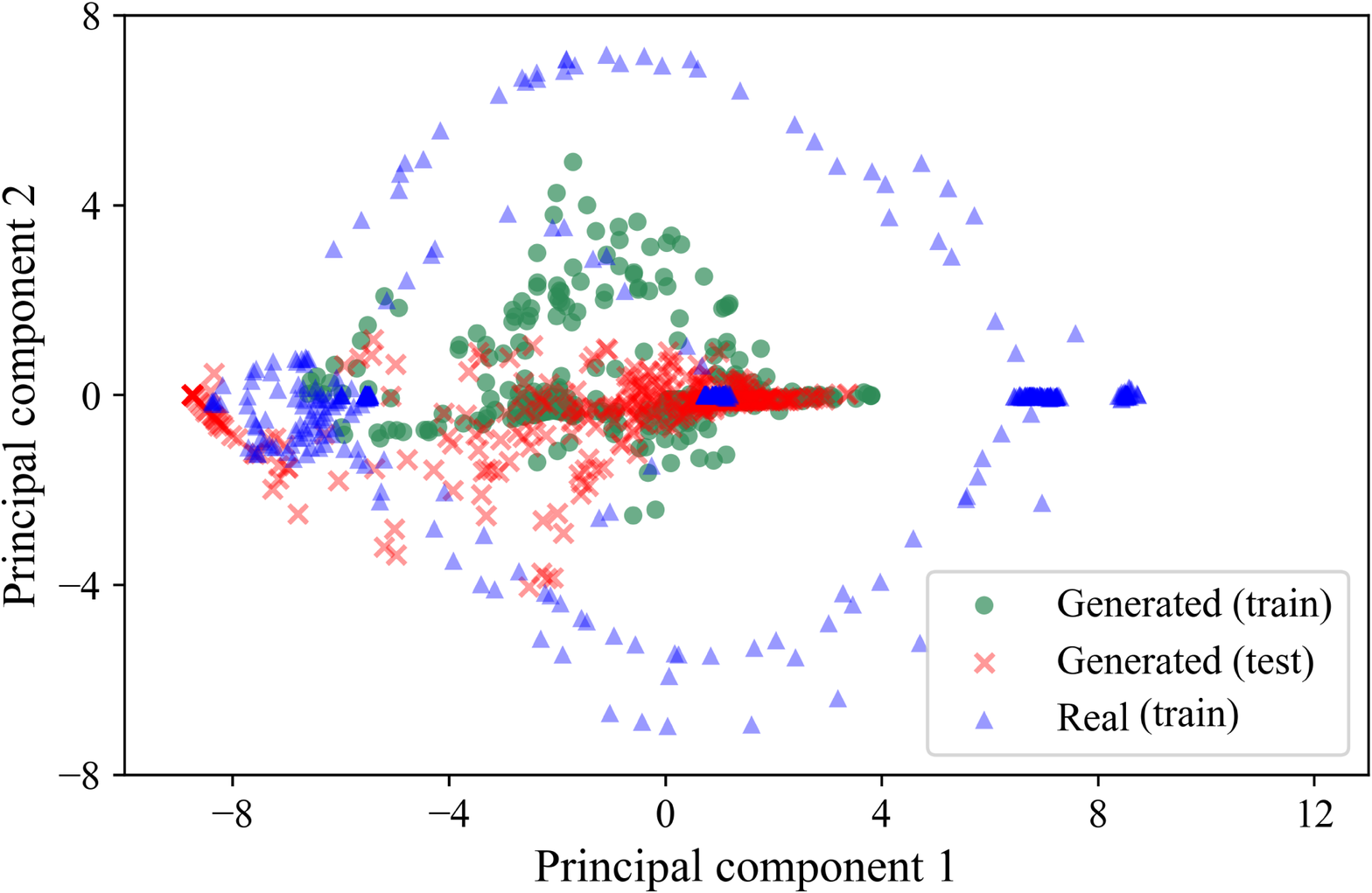}
\end{minipage}%
}%
\subfloat[]{
\begin{minipage}[t]{0.3\linewidth}
\centering
\includegraphics[width=5cm]{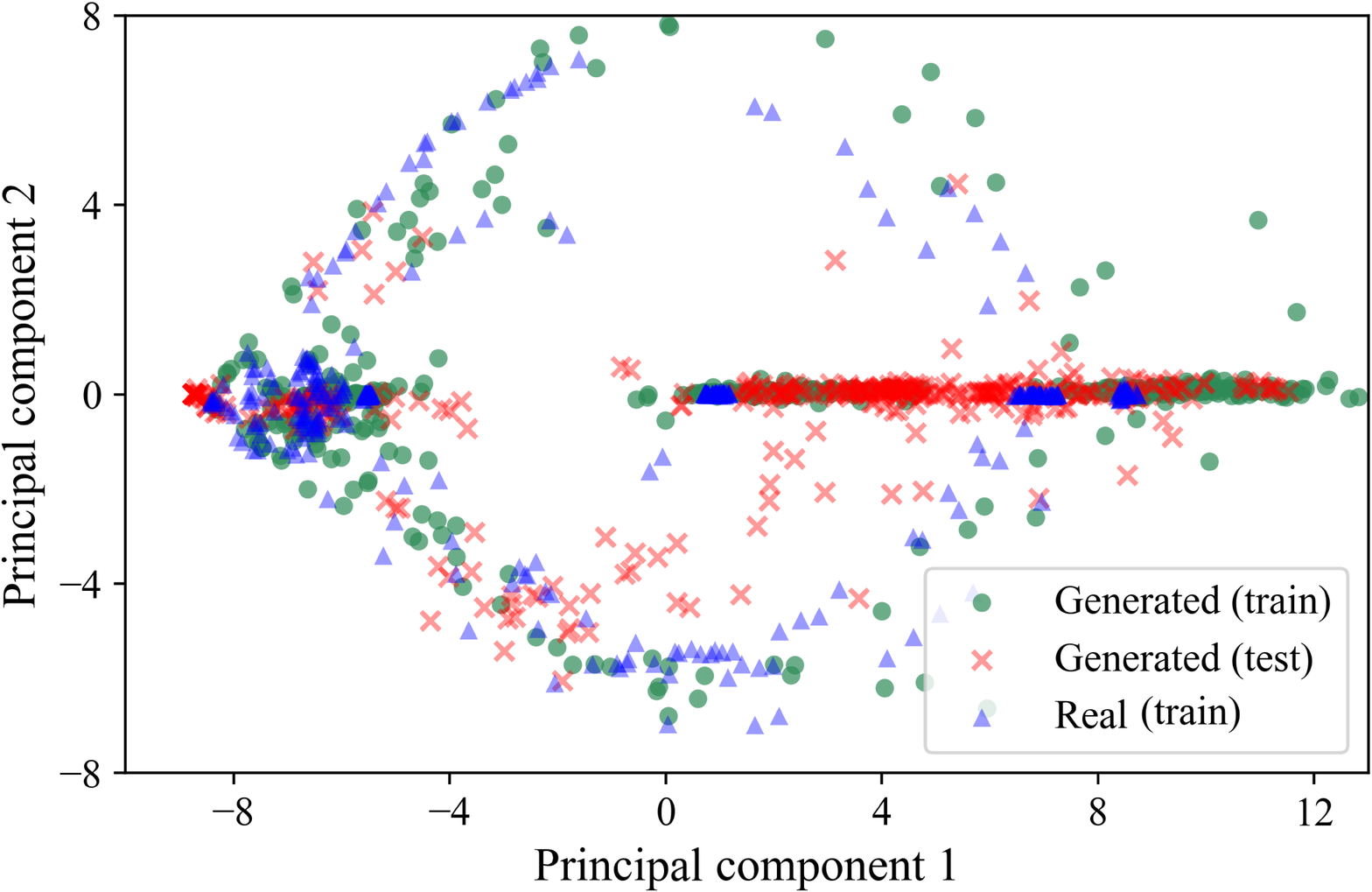}
\end{minipage}%
}%
\caption{Two-dimensional visualization of on-state samples for dishwashers in the REDD dataset with different models: (a) SGN, (b) SCANet without branch-wise gates, and (c) SCANet with branch-wise gates. We show 10\% of the samples of $\mathbf{y}$ (blue dots) in the training set and their corresponding $\hat{\mathbf{y}}$ (green dots). The red crosses correspond to $\hat{\mathbf{y}}$ for the test set. Four modes with high-density blue dots are identified in (a), each representing a power consumption level. Only SCANet with branch-wise gates can cover all the modes for $\hat{\mathbf{y}}$ in the test set.}
\label{GAN_mode}
\end{figure*}

We carry out an ablation study for the appliances in the REDD dataset and report the MAEs in Table \ref{ablation} (each MAE value is averaged over 3 trials). The first row corresponds to SGN. It is observed in the table that each component we add can reduce the MAE and that the components are mutually compatible, as lower MAEs can be achieved when they are combined. Specifically, the combination of the additional modules and the adversarial loss greatly improves the performance of the model. The improvement for fridge is mainly contributed by on-state augmentation, which helps the model adapt to a different power consumption level in the test data.

It is then of great interest to analyze the mechanism behind the accuracy boost when the additional modules are combined with adversarial loss. After some investigation, we find out that the adversarial loss helps the model capture the power consumption modes as expected, and that the branch-wise gates allow the model to avoid the mode collapse phenomenon (see \cite{srivastava2017veegan} for an introduction to mode collapse of GAN). The effect of the adversarial loss is demonstrated in Fig. \ref{GAN_mode} with dishwashers in the REDD dataset as an example. Specifically, we plot the first two principal components of the 64-time-step sequences of $\mathbf{y}$ (real samples) and $\hat{\mathbf{y}}$ (generated samples) after principal component analysis (PCA). Only complete on-state sequences are considered for simplicity and clarity. Four modes of $\mathbf{y}$ in the training set are identified, and it is expected that $\hat{\mathbf{y}}$ in the training set would have the same distribution. Although $\hat{\mathbf{y}}$ in the test set may not have exactly the same distribution as $\mathbf{y}$ in the training set (different dishwashers may have varied consumption levels), it would be problematic if the distributions differ too much. In Fig. \ref{GAN_mode} (a), the SGN model produces sequences close to modes 1 and 2, but fails to cover modes 3 and 4 (only a small fraction gets close to mode 3). By contrast, the complete SCANet model covers all the modes (Fig. \ref{GAN_mode} (c)). However, it is shown in Fig. \ref{GAN_mode} (b) that the SCANet fails to cover modes 3 and 4 when the branch-wise gates are removed. Note that the branch-wise gates are not specifically designed for avoiding the mode collapse phenomenon. Nevertheless, the empirical observation for dishwasher in the REDD dataset shows that the branch-wise gates facilitates the incorporation of the adversarial loss.


\subsection{Practicability Verification With Limited Data and Partial Ground Truth}

The aforementioned experiments are carried out with at least several weeks of data from multiple houses and measurements of power consumptions of individual appliances are available. The individual-appliance-level measurements, however, may be impractical to obtain. In order to verify the performance of the proposed SCANet model when there is no access to fully labelled datasets (i.e., datasets containing consumptions of individual appliances) with large time spans, a different setting that uses less data with partial ground truth is adopted. More specifically, the new setting has the following features:
\begin{itemize}
\item We use the training data of the REDD dataset and test with the test data of the UK-DALE dataset, which puts higher demands on the generalization ability of the models.
\item It is assumed that appliance-level power consumption signals are inaccessible, but the on/off states of the appliances being considered are labelled. Thus, only the ground truth of on/off states are available.
\item Only a small proportion of training data is used to train the models.
\end{itemize} 

As the ground truths for appliance-level consumptions are unavailable, we modify the structures of SGN and SCANet and keep only the on/off state classification sub-networks in the models. As a result, the outputs of the models only contain on/off state predictions. The data in the REDD dataset is downsampled by a factor of 2 to match the sampling frequency of the UK-DALE dataset (i.e., $s$ is 32 and $w$ is 200). The step sizes for the REDD and the UK-DALE datasets are 4 and 2, respectively. Other hyper-parameters of the models remain unchanged. The adversarial losses are not added as the consumption values are unknown. Specifically, the experiments for the three appliances in the REDD dataset are designed as follows:
\begin{itemize}
\item Fridge: Two proportions, namely, 5\% and 10\% of the training data are used (the first 5\% or 10\% of each section in the training data as there are multiple sections). The total time span for the training data is roughly 3 days for the proportion of 10\%. On-state augmentation with $e^-=-0.15$ and $e^+=0.15$ is used.
\item Dishwasher: 20\% of the training data is used, which contains only 2 events of usage. On-state augmentation with $e^-=-1$ and $e^+=1$ is used.
\item Microwave: for microwave, 20\% of the training data is used and 12 microwave usage events are included. On-state augmentation with $e^-=-1$ and $e^+=1$ is used. In addition to this setting, we also experiment with adding part of the test data into the training data to mimic the process of gradually improving the model with the help of additional partially labelled data from the household being tested (e.g., from user feedbacks). Specifically, the additional data is taken from the beginning of the test data and is not used for evaluation. For the additional data, on-stage augmentation is also added with $e^-=-0.2$ and $e^+=0.2$.
\end{itemize} 

For performance evaluation, we use the $F_1$-score which is defined as
\begin{equation}
F_1 = \frac{2PR}{P+R},
\end{equation}
where $P=\frac{\rm{TP}}{\rm{TP}+\rm{FP}}$ is the precision and $R=\frac{\rm{TP}}{\rm{TP}+\rm{FN}}$ is the recall of the predictions for all the time steps. TP, FP, and FN stand for true positive, false positive, and false negative, respectively. For the predicted on/off state probabilities, values lower than 0.5 are considered as off and values greater or equal to 0.5 are considered as on. 

\begin{figure}[!tb]
\centering
\includegraphics[width=8.5cm]{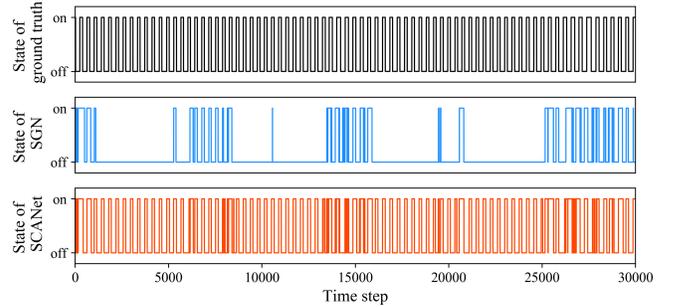}
\caption{Visualization of on/off states for fridge in the UK-DALE dataset. The models are trained using only 10\% of the training data from the REDD dataset.}
\label{fridge_UK}
\end{figure}

\begin{figure}[!tb]
\centering
\includegraphics[width=8.5cm]{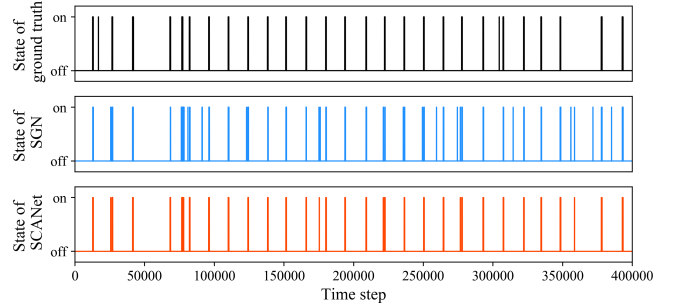}
\caption{Visualization of on/off states for dishwasher in the UK-DALE dataset. The models are trained using 20\% of the training data from the REDD dataset and only 2 on-state events are covered.}
\label{dishwasher_UK}
\end{figure}

\begin{figure}[!tb]
\centering
\includegraphics[width=8.5cm]{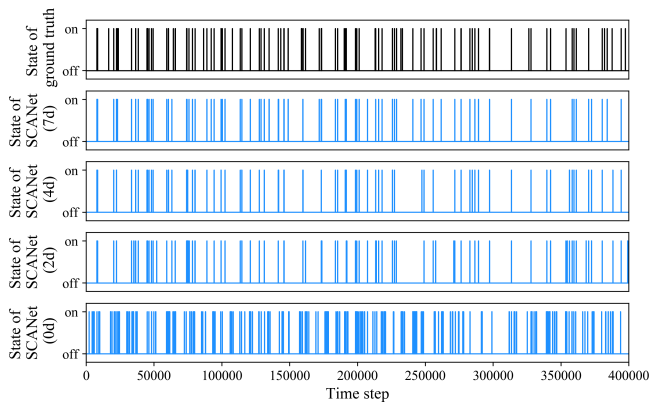}
\caption{Visualization of on/off states for microwave in the UK-DALE dataset. The models are trained with 20\% of the training data from the REDD dataset and some of the models have additional training data from the household being tested.}
\label{microwave_UK}
\end{figure}

\begin{table}[!tb]
\centering
\caption{Performance of the Models on the State Classification Task for Fridge}
\begin{tabular}{lrrrr}  
\toprule
Model      & Proportion &	Precision  	& Recall    & $F_1$-score    \\
\midrule
SGN        &  5\%       & 0.731         & 0.412     & 0.525         \\
SGN        &   10\%     & 0.724     	& 0.357     & 0.477         \\
SCANet     & 5\%		&0.812     		& 0.831     & 0.821         \\
SCANet     & 10\%		&\bf{0.823}     		& \bf{0.840}     & \bf{0.831}         \\
\bottomrule
\end{tabular}
\label{transfer_fridge}
\end{table}

\begin{table}[!tb]
\centering
\caption{Performance of the Models on the State Classification Task for Dishwasher With 20\% of Training Data}
\begin{tabular}{lrrr}  
\toprule
Model      & Precision  & Recall     & $F_1$-score    \\
\midrule
SGN                 & 0.588     & 0.550     & 0.567         \\
SCANet     & \bf{0.692}     & \bf{0.731}     & \bf{0.710}         \\
\bottomrule
\end{tabular}
\label{transfer_dishwasher}
\end{table}

\begin{table}[!tb]
\centering
\caption{Performance of the Models on the State Classification Task for Microwave With 20\% of Training Data}
\begin{tabular}{lrrrr}  
\toprule
Model               &  Added Days   & Precision  & Recall     & $F_1$-score    \\
\midrule
SGN                 &  0                 & 0.151      & 0.295      & 0.199         \\
SGN                 &  7                 & 0.171      & 0.452      & 0.249         \\
SCANet     &  0                 & 0.173      & 0.441      & 0.248         \\
SCANet     &  2                 & 0.637      & 0.531      & 0.579         \\
SCANet     &  4                 & 0.761      & 0.559      & 0.644         \\
SCANet     &  7                 & \bf{0.852}      & \bf{0.622}      & \bf{0.689}         \\
\bottomrule
\end{tabular}
\label{transfer_microwave}
\end{table}

The performance of the models for fridge is shown in Table \ref{transfer_fridge}, and the on/off states are visualized in Fig. \ref{fridge_UK}. It is clear in the results that the SCANet has higher recall than SGN, and the $F_1$-score of SCANet is much higher. Specifically, the SGN model predicts a lot of off states when the fridge is working. The results for dishwasher are shown in Table \ref{transfer_dishwasher} and Fig. \ref{dishwasher_UK}. With only two usage events in the training data, it is quite impressive that the SCANet model is able to have high precision and recall at the same time. For the time range of Fig. \ref{dishwasher_UK}, there are 29 usage events for the ground truth, and the SCANet model responses to 27 of them. Meanwhile, only 2 false positive cases are produced.

The results for microwave are shown in Table \ref{transfer_microwave} and Fig. \ref{microwave_UK}. When no data from the test dataset is added, the performances of both models are not satisfactory. This indicates that transferring a model learned on the REDD dataset to the UK-DALE dataset is problematic for microwave. The reason transferring the model of fridge is easier is that fridges generally have a unique cyclic power consumption pattern with a relatively low consumption level. The consumption pattern of dishwashers is also quite unique and the time span for a single usage is relatively long. The usage pattern of microwaves, however, may be confused with other appliances as it mainly consists of sparse, short windows with high power consumptions. As a result, adding partially labelled data from the test data greatly improves the performance of the SCANet model. When the data of a week containing 19 microwave usage events is added, the $F_1$-score of the model increases to 0.689. Further, if we consider the number of events recorded in Fig. \ref{microwave_UK}, the precision and recall are $63 / 65 \approx 0.969$ and $63 / 77 \approx 0.818$ for the model trained with data of 7 additional days in the test data.

In short, we have shown that the proposed SCANet model has a better performance than SGN in the new experiment setting. In addition, a model trained in this manner may also be used to facilitate unsupervised NILM approaches (e.g., help assign the disaggregation results to specific appliances). 

\section{Conclusion}

We develop a scale- and context-aware CNN model, namely SCANet, for the task of NILM in this paper. Experiment results show that the proposed SCANet significantly reduces the estimation error of the disaggregated appliance-level power consumption. Adding adversarial loss and on-state augmentation are proven to be useful for certain appliances. In addition to the comparisons with the state-of-the-art, we also provide some observations on the working mechanisms of the modules by diving into the intermediate network layers. We show that the scale- and context-aware modules are functioning as expected, which contribute to the improvement in disaggregation accuracy. 

In order for NILM techniques to function properly for real-world applications, an important path for future work is to combine the merits of supervised and unsupervised learning. One possibility is to combine the results from supervised and unsupervised models to produce better results. Another direction is to design a practical setting for semi-supervised learning and try to incorporate unlabelled or partially labelled data into the training process of a model.

\section*{Acknowledgments}

The authors thank Changho Shin for providing the pre-processed REDD dataset used in \cite{shin2018subtask}. We are also grateful for the support of NVIDIA Corporation with the donation of a Titan Xp GPU. This work was partially supported by the 2019 Seed Fund Award from CITRIS and the Banatao Institute at the University of California, and the Hellman Fellowship.

{
\small
\bibliographystyle{IEEEtran}%
\bibliography{IEEE.bib}
}

\begin{IEEEbiography} [{\includegraphics[width=1in,height=1.25in,clip,keepaspectratio]{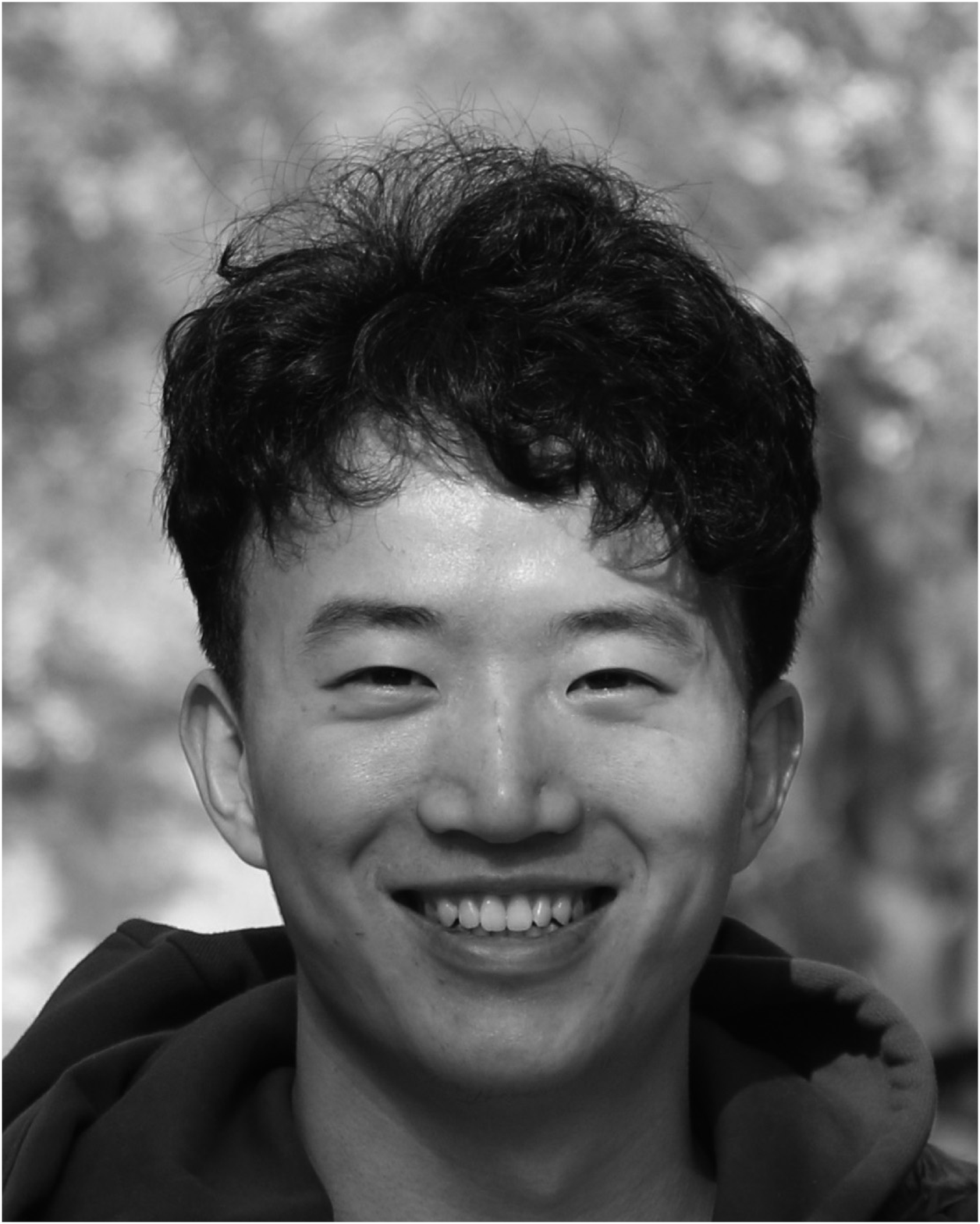}}]
{Kunjin Chen} received the B.Sc. degree in electrical engineering from Tsinghua University, Beijing, China, in 2015. Currently, he is a Ph.D. candidate with the Department of Electrical Engineering, Tsinghua University. 

His research interests include applications of machine learning and data science in power systems.
\end{IEEEbiography}

\vspace{-25pt}

\begin{IEEEbiography} [{\includegraphics[width=1in,height=1.25in,clip,keepaspectratio]{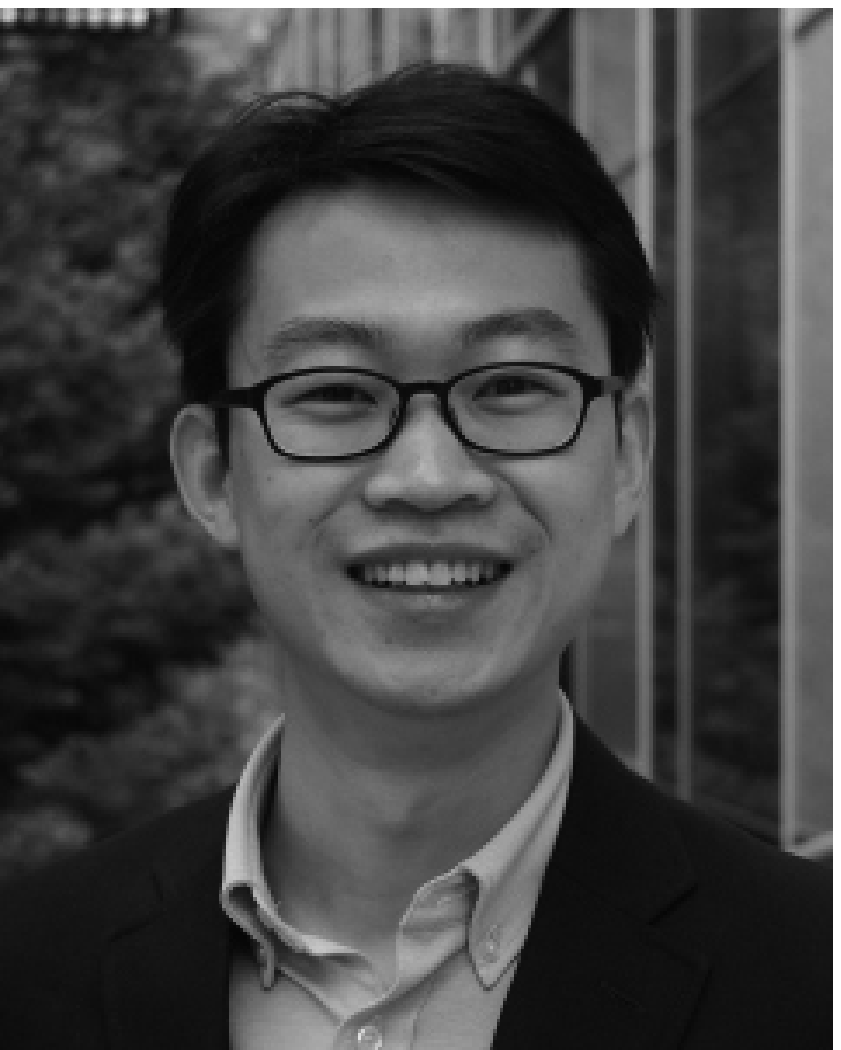}}]
{Yu Zhang} (M'15) received the Ph.D. degree in electrical and computer engineering from the University of Minnesota, Minneapolis, MN, USA, in 2015.

He is an Assistant Professor in the ECE Department of UC Santa Cruz. Prior to joining UCSC, he was a postdoc at UC Berkeley and Lawrence Berkeley National Laboratory. His research interests span the broad areas of cyber-physical systems, smart power grids, optimization theory, machine learning and big data analytics.
\end{IEEEbiography}

\vspace{-25pt}

\begin{IEEEbiography} [{\includegraphics[width=1in,height=1.25in,clip,keepaspectratio]{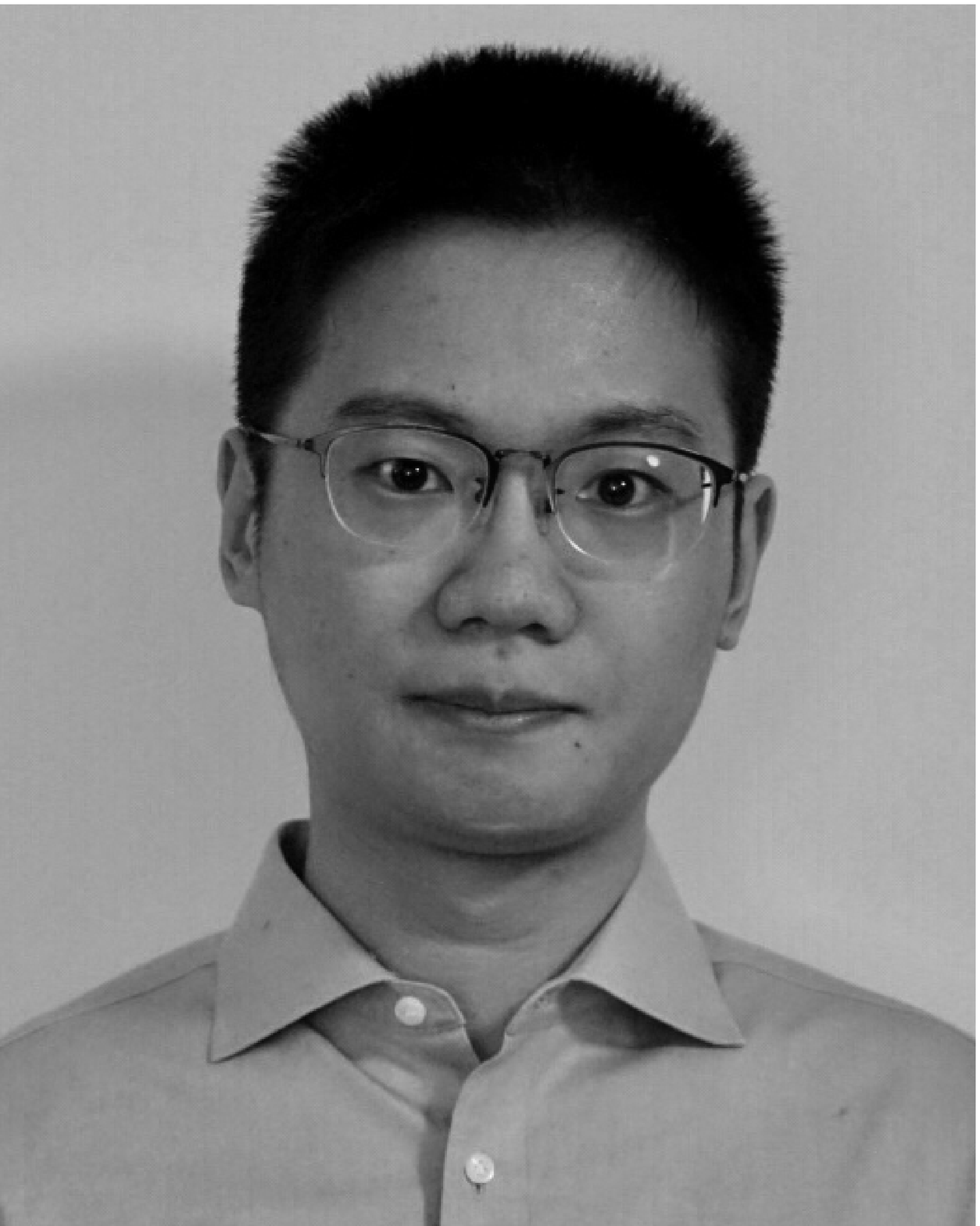}}]
{Qin Wang} received his B.Eng degree from Dept. Electrical Engineering, Tsinghua University in 2015 and Master's degree from Dept. Information Technology and Electrical Engineering, ETH Z\"urich in 2018. Currently, he is a Ph.D. student in the intelligent maintenance systems group at ETH Z\"urich. 

His research interests include weakly supervised learning and domain adaptation. 
\end{IEEEbiography}

\vspace{-25pt}

\begin{IEEEbiography} [{\includegraphics[width=1in,height=1.25in,clip,keepaspectratio]{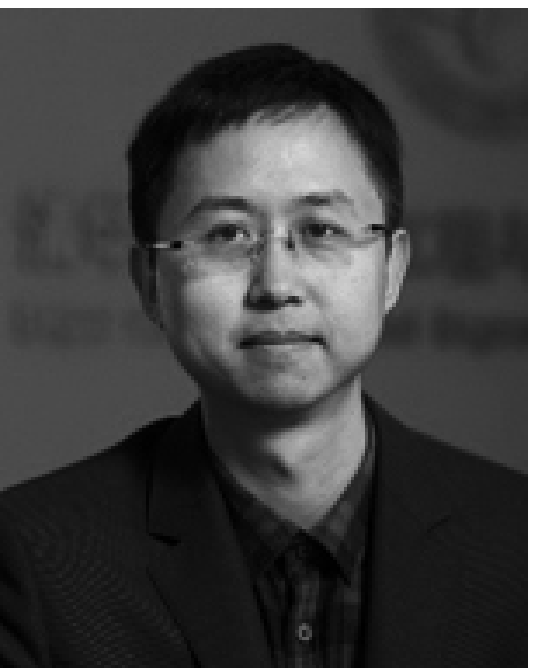}}]
{Jun Hu} (M'10) received his B.Sc., M.Sc., and Ph.D. degrees in electrical engineering from the Department of Electrical Engineering, Tsinghua University in Beijing, China, in July 1998, July 2000, July 2008. 

Currently, he is an associate professor in the same department. His research fields include overvoltage analysis in power system, sensors and big data, dielectric materials and surge arrester technology.
\end{IEEEbiography}

\vspace{-25pt}

\begin{IEEEbiography} [{\includegraphics[width=1in,height=1.25in,clip,keepaspectratio]{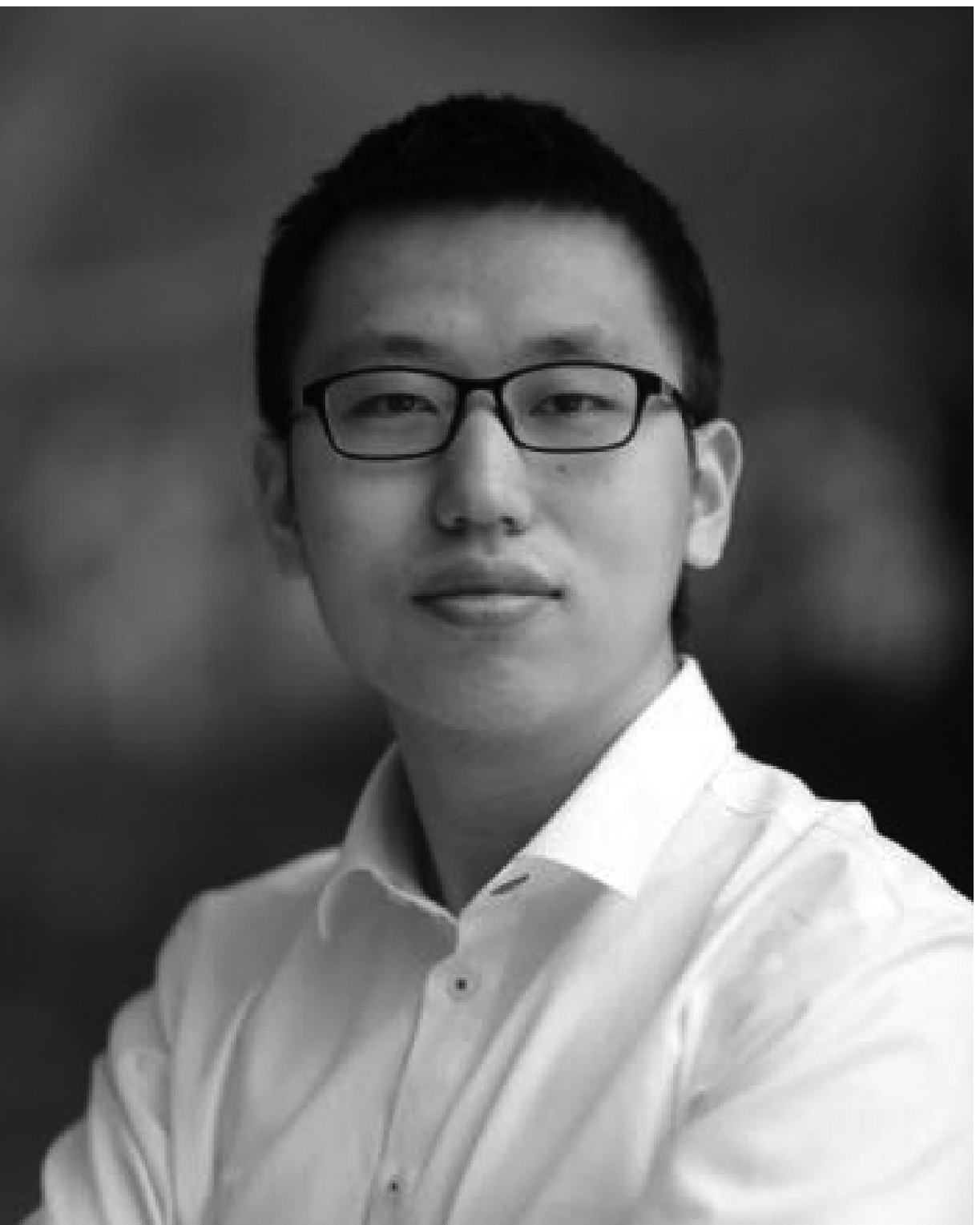}}]
{Hang Fan} received his B.Eng degree from Dept. Electrical Engineering, Sichuan University in 2015. He is currently a Ph.D. candidate with the Department of Electrical Engineering, Tsinghua University.

His research interests include the application of artificial intelligence in the field of power systems.
\end{IEEEbiography}

\vspace{-25pt}

\begin{IEEEbiography} [{\includegraphics[width=1in,height=1.25in,clip,keepaspectratio]{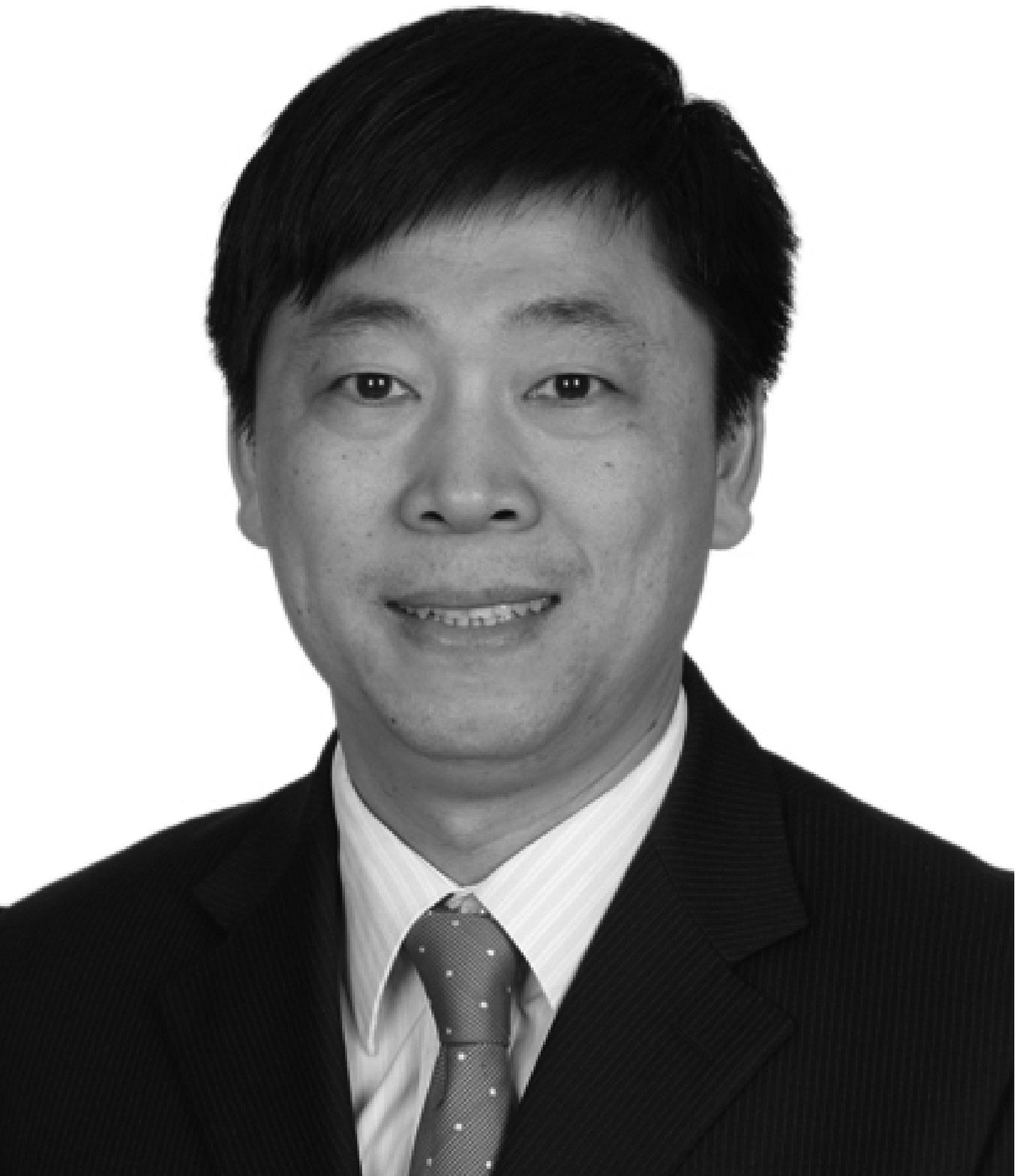}}]
{Jinliang He} (M'02--SM'02--F'08) received the B.Sc. degree from Wuhan University of Hydraulic and Electrical Engineering, Wuhan, China, the M.Sc. degree from Chongqing University, Chongqing, China, and the Ph.D. degree from Tsinghua University, Beijing, China, all in electrical engineering, in 1988, 1991 and 1994, respectively.

He became a Lecturer in 1994, and an Associate Professor in 1996, with the Department of Electrical Engineering, Tsinghua University. From 1997 to 1998, he was a Visiting Scientist with Korea Electrotechnology Research Institute, Changwon, South Korea, involved in research on metal oxide varistors and high voltage polymeric metal oxide surge arresters. From 2014 to 2015, he was a Visiting Professor with the Department of Electrical Engineering, Stanford University, Palo Alto, CA, USA. In 2001, he was promoted to a Professor with Tsinghua University. He is currently the Chair with High Voltage Research Institute, Tsinghua University. He has authored five books and 400 technical papers. His research interests include overvoltages and EMC in power systems and electronic systems, lightning protection, grounding technology, power apparatus, and dielectric material.
\end{IEEEbiography}

\end{document}